\RequirePackage{lineno} 
\documentclass[aps,prc,superscriptaddress,showpacs,floatfix,nofootinbib,twocolumn]{revtex4}
\usepackage{graphicx} 
\usepackage{float} 
\usepackage{amssymb}
\usepackage{url}
\usepackage{harpoon}
\usepackage{amsmath}
\usepackage{multirow}
\usepackage{CJK}
\usepackage[colorlinks,breaklinks]{hyperref}
\hypersetup{linkcolor=blue,citecolor=blue,filecolor=black,urlcolor=blue}
\newcommand{\pT} {\ensuremath{p_{\mathrm{T}}}}

\begin{document}
\begin{CJK*}{UTF8}{}
\title{Elucidating the event-by-event flow fluctuations in heavy-ion collisions via the event shape selection technique}
\newcommand{\sunysb}{Department of Chemistry, Stony Brook University, Stony Brook, NY 11794, USA}
\newcommand{\bnl}{Physics Department, Brookhaven National Laboratory, Upton, NY 11796, USA}
\newcommand{\columbia}{Current address: Columbia University and Nevis Laboratories, Irvington, NY 10533, USA}
\author{Peng Huo(\CJKfamily{gbsn}霍鹏)}\affiliation{\sunysb}
\author{Jiangyong Jia(\CJKfamily{gbsn}贾江涌)}\email[Correspond to\ ]{jjia@bnl.gov}
\affiliation{\sunysb}\affiliation{\bnl}
\author{Soumya Mohapatra}\affiliation{\sunysb}\affiliation{\columbia}
\begin{abstract}
The presence of large event-by-event flow fluctuations in heavy ion collisions at RHIC and the LHC provides an opportunity to study a broad class of flow observables. This paper explores the correlations among harmonic flow coefficients $v_n$ and their phases $\Phi_n$, as well as the rapidity fluctuation of $v_n$. The study is carried out using the Pb+Pb events generated by the AMPT model with fixed impact parameter. The overall ellipticity or triangularity of events is varied by selecting on the eccentricities $\epsilon_n$ or the magnitudes of the flow vector $q_n$ in a subevent for $n=2$ and 3, respectively. The responses of the harmonic coefficients, the event-plane correlations, and the rapidity fluctuations, to the change in $\epsilon_n$ and $q_n$ are then systematized. Strong positive correlations are observed among all even harmonics $v_2, v_4$, and $v_6$ (all increase with $q_2$), between $v_2$ and $v_5$ (both increase with $q_2$) and between $v_3$ and $v_5$ (both increase with $q_3$), consistent with the effects of non-linear collective response. In contrast, an anti-correlation is observed between $v_2$ and $v_3$ similar to that seen between $\epsilon_2$ and $\epsilon_3$. These correlation patterns are found to be independent of whether selecting on $q_n$ or $\epsilon_n$, validating the ability of $q_n$ in selecting the initial geometry. A forward/backward asymmetry of $v_n(\eta)$ is observed for events selected on $q_n$ but not on $\epsilon_n$, reflecting dynamical fluctuations exposed by the $q_n$ selection. Many event-plane correlators show good agreement between $q_n$ and $\epsilon_n$ selections, suggesting that their variations with $q_n$ are controlled by the change of $\epsilon_n$ in the initial geometry. Hence these correlators may serve as promising observables for disentangling the fluctuations generated in various stages of the evolution of the matter created in heavy ion collisions.
\end{abstract}
\pacs{25.75.Dw} \maketitle 
\end{CJK*}
\section{Introduction} 
High energy heavy ion collisions at RHIC and the LHC have created a new form of nuclear matter comprised of deconfined, yet strongly interacting quarks and gluons. This matter exhibits strong collective and anisotropic flow in the transverse plane, which is well described by relativistic hydrodynamics~\cite{Alver:2010dn,Gale:2012in,Heinz:2013th}. The magnitude of the anisotropic flow has been to found to be sensitive to the transport properties, as well as the space-momentum profile in the initial state. A central goal of current research is to understand the nature of various fluctuations in the initial state and how these fluctuations influence the hydrodynamic evolution of the matter in the final state.

In heavy ion collisions, the anisotropy of the particle distribution in azimuthal angle $\phi$ is customarily characterized by a Fourier series:
\begin{equation}
\label{eq:flow0}
\frac{dN}{d\phi}\propto1+2\sum_{n=1}^{\infty}v_{n}\cos n(\phi-\Phi_{n})\;,
\end{equation}
where $v_n$ and $\Phi_n$ (event plane or EP) represent the magnitude and phase of the $n^{\mathrm{th}}$-order harmonic flow. These flow harmonics have been associated to various shape components of the created matter~\cite{Alver:2010gr}. The magnitude and direction of each shape component can be estimated via a simple Glauber model from the transverse positions (r, $\phi$) of the participating nucleons~\cite{Qin:2010pf,Teaney:2010vd}:
\begin{eqnarray}
\label{eq:ena}
\epsilon_n &=& \frac{\sqrt{\langle r^m\cos n\phi\rangle^2+\langle r^m\sin n\phi\rangle^2}}{\langle r^m\rangle},\\\label{eq:glau} 
\tan (n\Phi_n^*+\pi)&=&\frac{\langle r^m\sin n\phi\rangle}{\langle r^m\cos n\phi\rangle},\\\nonumber
&&m=3\; {\rm if}\; n=1, m=n\; \rm{ for}\; n>1
\end{eqnarray} 
with $\epsilon_n$ and $\Phi^*_n$ referred to as the eccentricity and participant plane (PP), respectively. Model calculations suggest that hydrodynamic response to the shape component is linear for the first few flow harmonics, i.e. $\Phi_n\approx \Phi_n^*$ and $v_n\propto \epsilon_n$ for $n=$1--3~\cite{Teaney:2010vd,Qiu:2011iv}. But these simple relations are violated for higher-order harmonics, due to strong mode-mixing effects intrinsic in the collective expansion~\cite{Qiu:2011iv,Gardim:2011xv,Teaney:2012ke}.

The presence of large event-by-event (EbyE) fluctuations of the initial geometry suggests a general set of observables that involve correlations between $v_n$ and $\Phi_n$:
\begin{equation}
\label{eq:flow}
p(v_n,v_m,...., \Phi_n, \Phi_m, ....)=\frac{1}{N_{\mathrm{evts}}}\frac{dN_{\mathrm{evts}}}{dv_ndv_m...d\Phi_{n}d\Phi_{m}...},
\end{equation}
with each variable being a function of $\pT$, $\eta$ etc~\cite{Gardim:2012im}. Among these, the joint probability distribution of the EP angles:
\small{
\begin{eqnarray}
\nonumber
\frac{dN_{\mathrm{evts}}}{d\Phi_{1}d\Phi_{2}...d\Phi_{l}} &\propto& \sum_{c_n=-\infty}^{\infty} a_{c_1,c_2,...,c_l} \cos(c_1\Phi_1+c_2\Phi_2...+c_l\Phi_l),\\\label{eq:ep}
a_{c_1,c_2,...,c_l}&=&\left\langle\cos(c_1\Phi_1+c_2\Phi_2+...+c_l\Phi_l)\right\rangle
\end{eqnarray}}\normalsize
can be reduced to the following event-plane correlators required by symmetry~\cite{Bhalerao:2011yg,Qin:2011uw,Jia:2012ma}:
\begin{eqnarray}
\label{eq:ep2}
\left\langle\cos(c_1\Phi_1+2c_2\Phi_2...+lc_l\Phi_l)\right\rangle, c_1+2c_2...+lc_l=0.
\end{eqnarray}
These observables are sensitive to the fluctuations in the initial density profile and the final state hydrodynamics response~\cite{Teaney:2012ke}.

Earlier flow measurements were aimed at studying the individual $v_n$ coefficients for $n=$1--6 averaged over many events~\cite{Adare:2011tg,star:2013wf,Aamodt:2011by,Aad:2012bu,CMS:2012wg}. Recently, the LHC experiments exploited the EbyE observables defined in Eq.~\ref{eq:flow} by performing the first measurement of $p(v_n)$~\cite{Aad:2013xma} for $n=2-4$ and fourteen correlators involving two or three event planes~\cite{Jia:2012sa,ALICE:2011ab}. The measured event-plane correlators are reproduced by EbyE hydrodynamics~\cite{Qiu:2012uy,Teaney:2012gu} and AMPT transport model~\cite{Bhalerao:2013ina} calculations. The EP correlation measurement provides detailed insights on the non-linear hydrodynamic response, for example the correlators $\left\langle\cos 4(\Phi_{2}-\Phi_{4})\right\rangle$ and $\left\langle\cos 6(\Phi_{3}-\Phi_{6})\right\rangle$ mainly arise from the non-linear effects, which couple $v_4$ to $(v_2)^2$ and $v_6$ to $(v_3)^2$. Similarly, the correlator $\left\langle\cos (2\Phi_{2}+3\Phi_{3}-5\Phi_5)\right\rangle$ is driven by the coupling between $v_5$ and $v_2v_3$~\cite{Gardim:2011xv,Teaney:2012ke}.

This paper focuses on two subsets of the observables defined by Eq.~\ref{eq:flow}: $p(v_n,v_m)$ and $p(v_n, \Phi_m, \Phi_l, ....)$, which can provide further insights on the linear and non-linear effects in the hydrodynamics response. The correlation $p(v_n,v_m)$  quantifies directly the coupling between $v_m$ and $v_n$, while $p(v_n, \Phi_m, \Phi_l, ...)$ allows us to study how the event-plane correlations couples to a specific flow harmonics $v_n$. The probability distributions of these correlations are difficult to measure directly, instead we explore them systematically using the recently proposed event shape selection method~\cite{Schukraft:2012ah} (also investigated in Ref.~\cite{Petersen:2013vca,Lacey:2013eia}): Events in a given centrality interval are first classified according to the observed $v_n$ signal in certain $\eta$ range, and the $p(v_m)$ and $p(\Phi_m, \Phi_l, ...)$ are then measured in other $\eta$ range for each class. The event shape observables should be those that correlate well with the $\epsilon_n$ of the initial geometry, such as the observed $v_1$ (dipolar flow), $v_2$ and $v_3$. The roles of these selection variables are similar to the event centrality, except that they further divide events within the same centrality class.

The event shape selection method also provides a unique opportunity to investigate the longitudinal dynamics of the collective flow. For example, events selected with large $v_2$ in one pseudorapidity window, in addition to having bigger $\epsilon_2$, may also have stronger density fluctuations, larger initial flow or smaller viscous correction~\cite{Pang:2012he}. Studying how the $v_n$ values or EP correlations vary with the $\eta$ separation from the selection window may provide better insights on the longitudinal dynamics in the initial and the final states. Earlier efforts in this front can be found in Refs.~\cite{Petersen:2011fp,Pang:2012he,Xiao:2012uw}.

In this paper, we apply the event shape selection technique to events generated by the AMPT model, to investigate the $p(v_n,v_m)$, $p(v_n, \Phi_m, \Phi_l, ....)$, and the longitudinal flow fluctuations. These correlations are studied for events binned according to the observed $v_2/v_3$ signal, which are then compared with results for events binned directly in $\epsilon_2/\epsilon_3$. This comparison helps to elucidate whether the changes in the correlation are driven mostly by the selection of the initial geometry or due to additional dynamics in the final state. This study also help to develop and validate the analysis method to be used in the actual data analysis.

The structure of the paper is as follows: Section~\ref{sec:1} introduces the observables and method of the event shape selection in the AMPT model. Section~\ref{sec:2} studies how the correlations among the eccentricities and PP angles vary with event shape selection. Section~\ref{sec:3} presents a study of the rapidity fluctuations of flow. Section~\ref{sec:4} studies how the correlations among the $v_n$'s and $\Phi_n$'s vary with event shape selection. Section~\ref{sec:5} gives a discussion and summary of the results.

\section{The method}
\label{sec:1}
A Muti-Phase Transport model (AMPT)~\cite{Lin:2004en} has been used frequently to study the higher-order $v_n$ associated with $\epsilon_n$ in the initial geometry~\cite{Xu:2011jm,Xu:2011fe,Ma:2010dv}. It combines the initial fluctuating geometry based on Glauber model from HIJING with the final state interaction via a parton and hadron transport model. The collective flow in this model is driven mainly by the parton transport. The AMPT simulation in this paper is performed with string-melting mode with a total partonic cross-section of 1.5 mb and strong coupling constant of $\alpha_s=0.33$~\cite{Xu:2011fe}, which has been shown to reproduce reasonably the $\pT$ spectra and $v_n$ data at RHIC and LHC~\cite{Xu:2011fe,Xu:2011fi}. The initial condition of the AMPT model with string melting has been shown to contain significant longitudinal fluctuations that can influence the collective dynamics~\cite{Pang:2012he,Pang:2012uw}.

\begin{figure}[h!]
\centering
\includegraphics[width=0.8\linewidth]{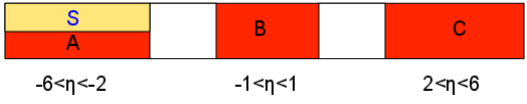}
\caption{\label{fig:m1} (Color online) The $\eta$ range of the subevent for the event shape selection (S) and three other subevents for correlation analysis (A, B and C). Note that the particles in $-6<\eta<-2$ are divided randomly and equally into subevents A and S.}
\end{figure}

The AMPT sample used in this study is generated for $b=8$~fm Pb+Pb collisions at LHC energy of $\sqrt{s_{NN}}=2.76$ TeV, corresponding to $\sim 30\%$ centrality. The particles in each event are divided into various subevents along $\eta$, one example division scheme is shown in Fig.~\ref{fig:m1}. Four subevents labelled as S, A, B, C, with at least 1 unit $\eta$ gap between any pair except between S and A, are used in the analysis. Note that particles in $-6<\eta<-2$ are divided randomly into two equal halves, labelled as S and A, respectively. The particles in subevent S are used only for the event shape selection purpose, and they are excluded for $v_n$ and event-plane correlation analysis. This choice of subevents and analysis scheme ensure that the event shape selection does not introduce non-physical correlations between S and A, B or C.

The flow vector in each subevent is calculated as:
\begin{eqnarray}
\nonumber
&&\overrightharp{q}_n =(q_{x,n},q_{y,n}) = \frac{1}{\Sigma_i w_i}\left(\textstyle\Sigma_i (w_i\cos n\phi_i), \Sigma_i (w_i\sin n\phi_i)\right)\;, \\\label{eq:me1}
&&\tan n\Psi_n = \frac{q_{y,n}}{q_{x,n}}\;,
\end{eqnarray}
where the weight $w_i$ is chosen as the $\pT$ of $i$-th particle and $\Psi_n$ is the measured event plane. Due to finite number effects, $\Psi_n$  smears around the true event-plane angle $\Phi_n$. Hence $q_n$ represents the weighted raw flow coefficients $v_n^{\mathrm{obs}}$, $q_n=\Sigma_i \left(w_i (v_n^{\mathrm{obs}})_i\right)/\Sigma_i w_i$. In this study, each subevent in Fig.~\ref{fig:m1} has 1400-3000 particles, so $q_n$ is expected to follow closely the true $v_n$.

For each generated event, the following quantities are calculated for $n=$1--6: $(\epsilon_n, \Phi^*_n)$ from initial state, $(q_n^{\mathrm{A}}, \Psi^{\mathrm{A}}_n)$ for subevent A, $(q_n^{\mathrm{B}}, \Psi^{\mathrm{B}}_n)$ for subevent B, $(q_n^{\mathrm{C}}, \Psi^{\mathrm{C}}_n)$ for subevent C, and $(q_n^{\mathrm{S}}, \Psi^{\mathrm{S}}_n)$ for subevent S, a total of 60 quantities. The event shape selection is performed by dividing the generated events into 10 bins in $q_2^{\mathrm{S}}$ or $q_3^{\mathrm{S}}$ with equal statistics. Similar event shape selection procedure is also performed by slicing the values of $\epsilon_2$ or $\epsilon_3$ directly, with the aim of studying how well the physics for events selected in the final state correlates with those selected purely on the initial geometry.

\begin{figure}[h!]
\centering
\includegraphics[width=1\linewidth]{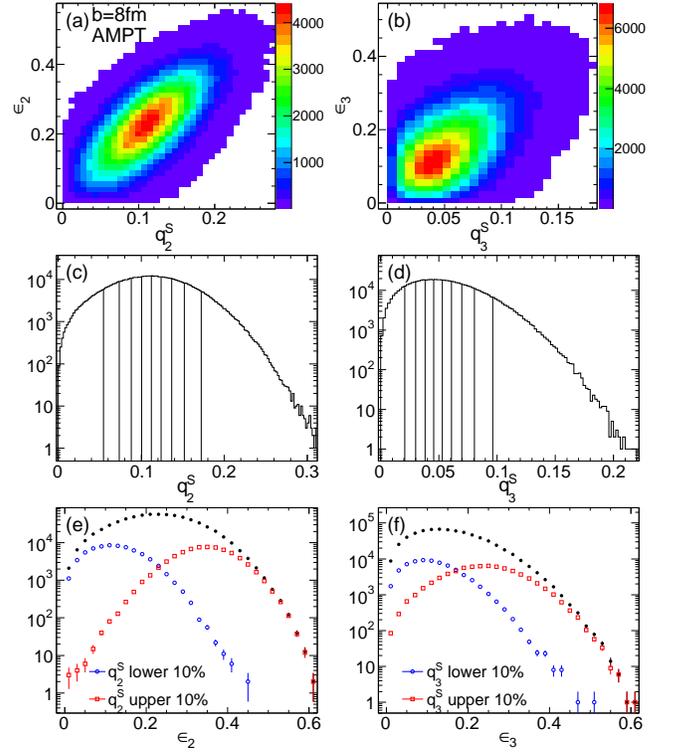}
\caption{\label{fig:m2} (Color online) Correlations between $\epsilon_n$ and magnitude of flow vector $q_n^{\mathrm{S}}$ calculated using half of the particles in $-6<\eta<-2$ (top panels), the 10 bins in $q_n^{\mathrm{S}}$ with equal statistics (middle panels) and the corresponding distributions of $\epsilon_n$ for events in the top 10\%, bottom 10\% and total of $q_n^{\mathrm{S}}$ (bottom panels). The results are calculated for $b=8$~fm for $n=2$ and $n=3$, and are shown in the left and right column, respectively.}
\end{figure}

Figure~\ref{fig:m2} shows the performance of the event shape selection on $q_2^{\mathrm{S}}$ and $q_3^{\mathrm{S}}$. Strong positive correlations between $\epsilon_n$ and $q_n^{\mathrm{S}}$ seen in the top panels reflect the fact that collective response is linear for $n=2$ and 3~\cite{Qiu:2012uy}. The bottom panels show that events selected with top 10\% of the $q_2^{\mathrm{S}}$ have a $\left\langle\epsilon_2\right\rangle$ value that is nearly 3 times that for events with the lower 10\% of $q_2^{\mathrm{S}}$. For $n=3$ the difference in $\epsilon_3$ in the two event classes is about a factor of 2. These results suggest that the ellipticity and triangularity of the initial geometry can be selected precisely by slicing the flow vector in the final state. 

In the event shape selection method, $p(v_m,v_n)$ is not directly calculated. Instead, the calculated correlation is:
\begin{eqnarray}
\label{eq:m2}
p(q_m^{\mathrm{S}},v_n^{\mathrm{obs}}) =  p(q_m^{\mathrm{S}})\times p(v_n^{\mathrm{obs}})_{q_m^{\mathrm{S}}},\;\; m=2,3
\end{eqnarray}
where conditional probability $p(v_n^{\mathrm{obs}})_{q_m^{\mathrm{S}}}$ represents the distribution of $v_n^{\mathrm{obs}}$ for events selected with given $q_m^{\mathrm{S}}$ value. To minimize non-flow effects, the $v_n^{\mathrm{obs}}$ is calculated for particles separated in $\eta$ from the subevent that provides the event plane. To minimize non-flow effects, a $\eta$ gap from the corresponding event plane in each case is required. The probability $p(v_n)_{q_m^{\mathrm{S}}}$ can be obtained from $p(v_n^{\mathrm{obs}})_{q_m^{\mathrm{S}}}$ via the unfolding technique~\cite{Aad:2013xma,Jia:2013tja}, or if one is interested in the event-averaged $v_n$ values, the standard method~\cite{Poskanzer:1998yz} can be applied for each $q_m^{\mathrm{S}}$ bin:
\begin{eqnarray}
\label{eq:m3}
v_n(\pT,\eta)_{q_m^{\mathrm{S}}} = \left[\frac{v_n^{\mathrm{obs}}(\pT,\eta)}{\mathrm{Res}\{ n\Psi_n \} }\right]_{q_m^{\mathrm{S}}},
\end{eqnarray}
where the event-plane resolution factor $\mathrm{Res}\{ n\Psi_n \}$ is calculated separately for A, B, and C via the three-subevent method, providing three independent $v_n$ estimates~\cite{Poskanzer:1998yz}. Since the magnitude and direction of the flow vector are uncorrelated, the event shape selection is not expected to introduce biases to the resolution correction. One special case of Eq.~\ref{eq:m3} is $n=m$, which probes into the rapidity fluctuation of the $v_n$ itself (see Section~\ref{sec:3}).

To calculate the event-plane correlation for each $q_m^{\mathrm{S}}$ bin, the standard method introduced by the ATLAS collaboration based on event-plane correlation~\cite{Jia:2012sa,Jia:2012ma}, and the method based on scalar products in Refs.~\cite{Luzum:2012da,Bhalerao:2013ina} are adopted:
\begin{eqnarray} 
\nonumber
\langle\cos (\Sigma \Phi)\rangle &=& \frac{\langle\cos (\Sigma \Psi)\rangle} {\mathrm{Res}\{c_1\Psi_1\}\mathrm{Res}\{c_22\Psi_2\}...\mathrm{Res}\{c_ll\Psi_l\}}\\\label{eq:m4a}
\\\nonumber
\langle\cos (\Sigma \Phi)\rangle_w &=& \frac{\langle q_1^{c_1} q_2^{c_2}... q_l^{c_l}\cos (\Sigma \Psi)\rangle} {\mathrm{Res}\{c_1\Psi_1\}_w\mathrm{Res}\{c_22\Psi_2\}_w...\mathrm{Res}\{c_ll\Psi_l\}_w}\\\label{eq:m4}
\end{eqnarray}
where shorthand notions $\Sigma \Phi = c_1\Phi_{1}+2c_2\Phi_{2}+...+lc_l\Phi_{l}$ and $\Sigma \Psi = c_1\Psi_{1}+2c_2\Psi_{2}+...+lc_l\Psi_{l}$ are used. They are referred to as the EP method (Eq.~\ref{eq:m4a}) and the SP method (Eq.~\ref{eq:m4}) for the rest of this paper. The resolution factors $\mathrm{Res}\{c_nn\Psi_n\}$ and $\mathrm{Res}\{c_nn\Psi_n\}_w$ are calculated via three-subevent method involving subevents A, B and C:
\small{
 \begin{eqnarray}
 \label{eq:m5a}
 &&{\mathrm{Res}}\{jn\Psi^{\mathrm A}_{n}\}=\sqrt{\frac{\left\langle {\cos\Delta\Psi^{AB}_n} \right\rangle\left\langle {\cos \Delta\Psi^{AC}_n}\right\rangle}{\left\langle {\cos \Delta\Psi^{BC}_n} \right\rangle}}.\\\nonumber
&&{\mathrm{Res}}\{jn\Psi^{\mathrm A}_{n}\}_w= \sqrt{\frac{\left\langle { (q_n^{\mathrm A}q_n^{\mathrm B})^j \cos \Delta\Psi^{AB}_n} \right\rangle \left\langle { (q_n^{\mathrm A}q_n^{\mathrm C})^j\cos \Delta\Psi^{AC}_n }\right\rangle}{\left\langle { (q_n^{\mathrm B}q_n^{\mathrm C})^j\cos \Delta\Psi^{BC}_n} \right\rangle}}.\\\label{eq:m5}
 \end{eqnarray}}\normalsize
where $\Delta\Psi^{AB}_n = jn \left(\Psi_n^{\mathrm A} - \Psi_n^{\mathrm B}\right)$ etc. Each $\Psi_n$ angle in Eq.~\ref{eq:m4} is calculated in a separate subevent to avoid auto-correlations. The two subevents involved in two-plane correlation are chosen as A and C in Fig.~\ref{fig:m1}, while the three subevents in three-plane correlation are chosen as A, B, and C in Fig.~\ref{fig:m1}. Note that selecting on $q_n^{\mathrm{S}}$ explicitly breaks the symmetry between subevents A and C even though they still have symmetric $\eta$ acceptance. Thus their resolution factors are different and need to be calculated separately.

\section{Correlations in the initial state}
\label{sec:2}
\begin{figure}[h!]
\centering
\includegraphics[width=0.9\linewidth]{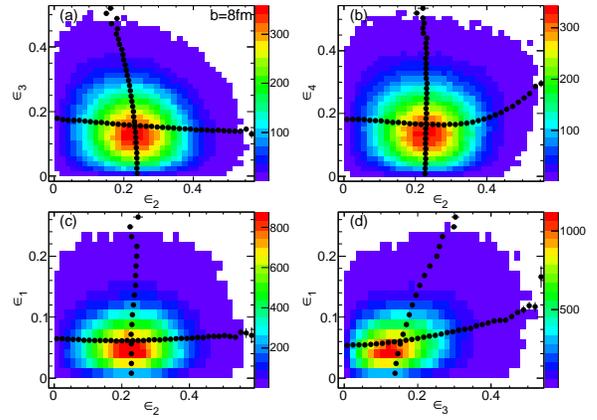}
\caption{\label{fig:i1} (Color online)  Selected correlations between $\epsilon_n$ of different order for Pb+Pb events at $b=8$~fm. More examples are given in Appendix~\ref{sec:7}. The $x$- and $y$-profiles of the 2D correlations are represented by the solid symbols.}
\end{figure}

Before discussing correlations in the final state, it is instructive to look first at how the initial geometry variables $\epsilon_n, \Phi^*_n$ and their correlations vary with the event shape selection. Figure~\ref{fig:i1} shows the correlations between pairs of $\epsilon_n$ for $n\leq4$ for the generated AMPT events. Significant correlations are observed between $\epsilon_2$ and $\epsilon_3$~\cite{Lacey:2013eia,ATLAS2014-022}, $\epsilon_1$ and $\epsilon_3$. The correlations between $\epsilon_1$ and $\epsilon_2$ are weak for this impact parameter but become more significant for $b=10$~fm (see Appendix~\ref{sec:7}). Since the hydrodynamic response is nearly linear for $n=1-3$~\cite{Qiu:2011iv}, these correlations are expected to survive into correlations between $v_n$ of respective order. The $\epsilon_2$ and $\epsilon_4$ correlation is also significant, especially for large $\epsilon_2$ values, this correlation may survive to the final state but it competes with non-linear effects expected for $v_4$~\cite{Gardim:2011xv}. 

\begin{figure}[h!]
\centering
\includegraphics[width=0.9\linewidth]{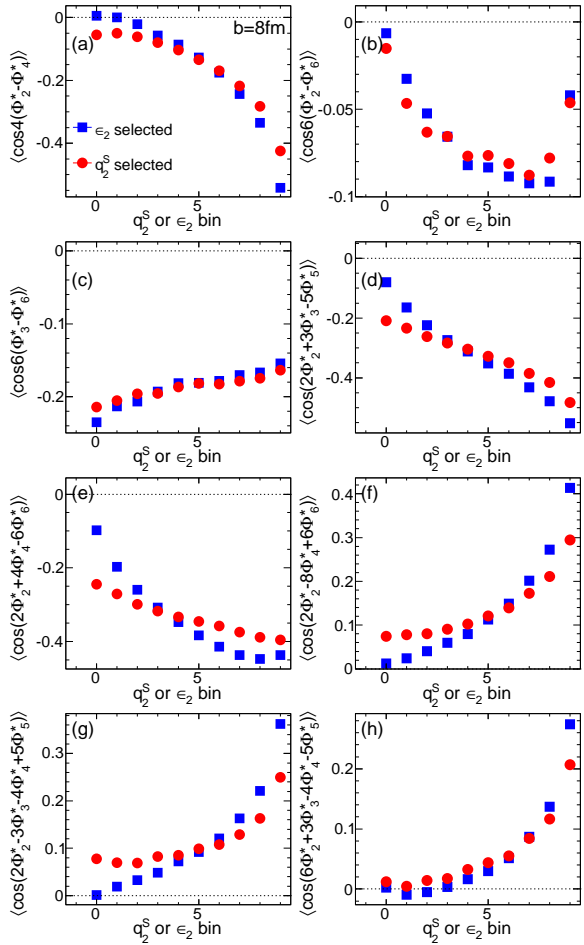}
\caption{\label{fig:i2} (Color online) Dependence of the participant-plane correlations on $\epsilon_2$ (boxes) or $q_2^{\mathrm{S}}$ (circles), calculated for Pb+Pb events with $b=8$~fm. Results for three two-plane, three three-plane and two four-plane correlators proposed in Refs.~\cite{Jia:2012ma,Jia:2012ju} are shown.}
\end{figure}

Figure~\ref{fig:i2} shows selected correlations between $\Phi_n^*$ of different order for events binned in $\epsilon_2$ (boxes) or $q_2^{\mathrm{S}}$ (circles)~\cite{Jia:2012ma,Jia:2012ju}. It is clear that the correlation signal varies dramatically with $\epsilon_2$, implying that the correlations between $\Phi_n^*$'s can vary a lot for events with the same impact parameter. Figure~\ref{fig:i2} also shows that events with different correlations in the initial geometry can be selected with nearly the same precision between using $q_2^{\mathrm{S}}$ and using $\epsilon_2$.

\section{The correlation between $v_n$ and $v_m$ and longitudinal fluctuations}
\label{sec:3}
\begin{figure}[h!]
\centering
\includegraphics[width=1\linewidth]{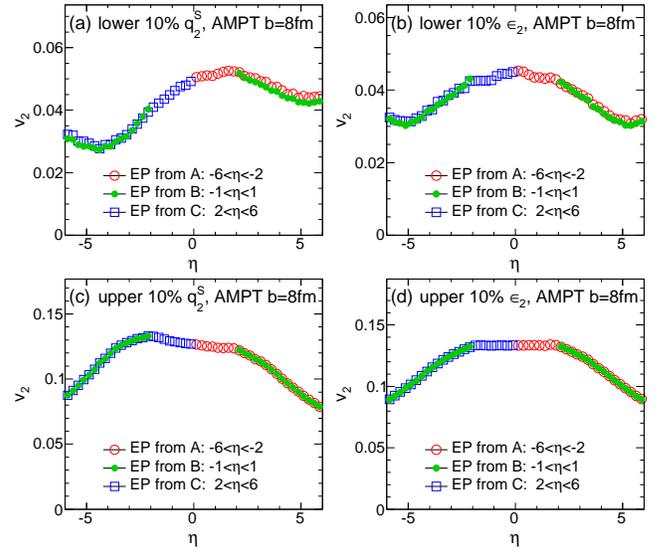}
\caption{\label{fig:eta1} (Color online) $v_2(\eta)$ for events selected with lower 10\% (top panels) and upper 10\% (bottom panels) of the values of either $q_2^{\mathrm{S}}$ (left panels) or $\epsilon_2$ (right panels) for AMPT Pb+Pb events with $b=8$~fm. In each case, the integral $v_2$ calculated for particles in $0.1<\pT<5$ GeV relative to the event plane of subevent A, B and C (Their $\eta$ coverages are indicated in the legend) with a minimum $\eta$ gap of 1 unit are shown.}
\end{figure}

Figure~\ref{fig:eta1} shows the $v_2(\eta)$ values for events selected for lower 10\% (top panels) and upper 10\% (bottom panels) of the values of either $q_2^{\mathrm{S}}$ (left panels) or $\epsilon_2$ (right panels). They are calculated via Eqs.~\ref{eq:m3} and \ref{eq:m5a} using all final state particles with $0.1<\pT<5$ GeV, excluding those particles used in the event shape selection (i.e. subevent S). The event-plane angles are calculated separately for the three subevents A, B, and C, and a minimum 1--2 unit of $\eta$ gap is required between $v_n(\eta)$ and the subevent used to calculate the event plane. Specifically, the $v_2$ values in $-6<\eta<0$ are obtained using the EP angle in subevent C covering $2<\eta<6$ (open boxes), the $v_2$ values in $0<\eta<6$ are obtained using the EP angle in subevent A covering $-6<\eta<-2$ (open circles), and the $v_2$ values in $|\eta|>2$ are also obtained using the EP angle in subevent B covering $|\eta|<1$ (solid circles).

\begin{figure*}[!t]
\centering
\includegraphics[width=1\linewidth]{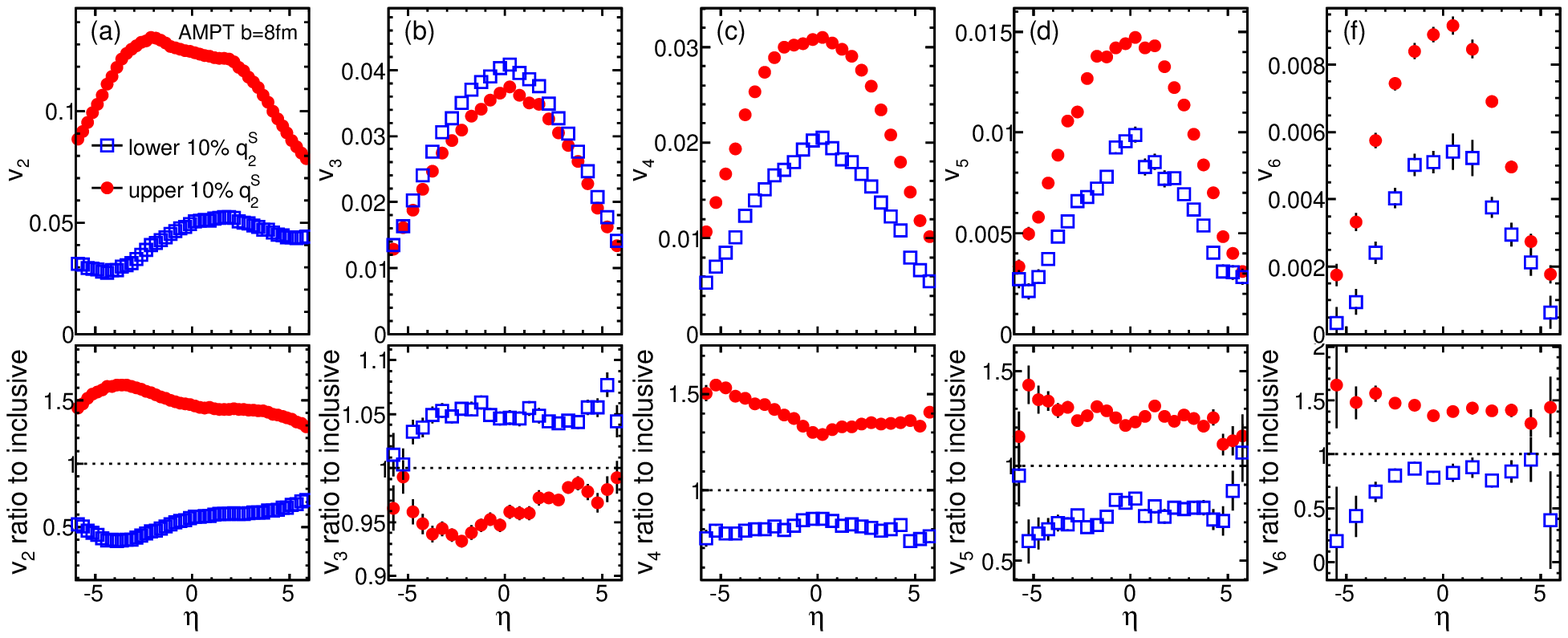}
\caption{\label{fig:eta2} (Color online) $v_n(\eta)$ for events selected with lower 10\% (open symbols) and upper 10\% (solid symbols) of the values of $q_2^{\mathrm{S}}$ for AMPT Pb+Pb events with $b=8$~fm. Results are shown for $v_2(\eta)$, $v_3(\eta)$,..., and $v_6(\eta)$ from left panel to the right panel. The ratios of $v_n(\eta)$ between events with $q_2^{\mathrm{S}}$ selection to the inclusive events are shown in the bottom panels.}
\end{figure*}
\begin{figure*}[!t]
\centering
\includegraphics[width=1\linewidth]{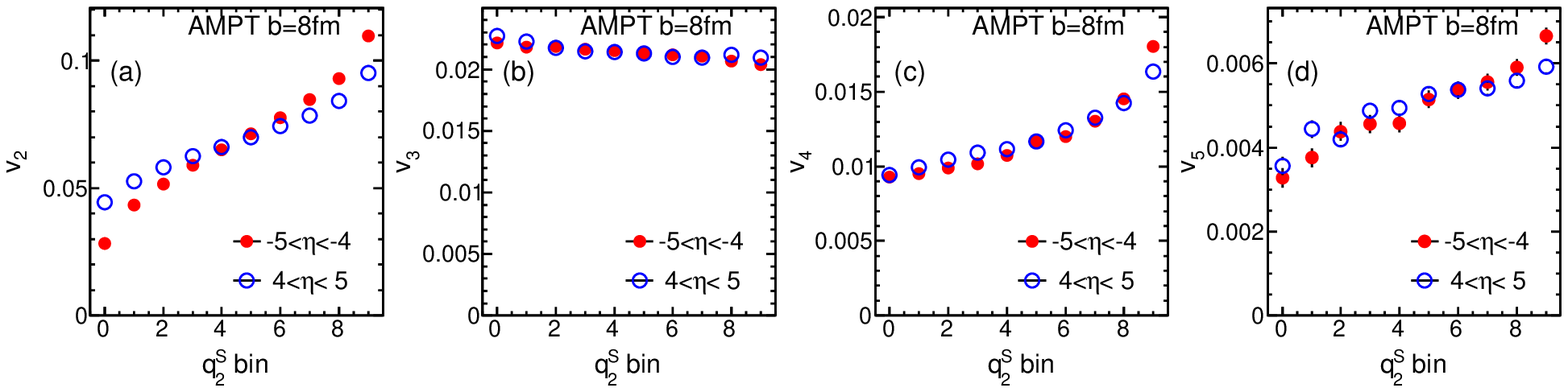}
\caption{\label{fig:eta3} (Color online)  Dependence of the $v_n$ on $q_2^{\mathrm{S}}$ in a forward ($-5<\eta<-4$) and a backward ($4<\eta<5$) pseudorapidity ranges.}
\end{figure*}

There are several interesting features in the observed $\eta$ dependence of $v_2$. The $v_2(\eta)$ values for events selected with lower 10\% values of $q_2^{\mathrm{S}}$ or $\epsilon_2$ are significantly lower (by a factor of 3) than for events selected with upper 10\%, indicating that the $v_2$ signal correlates well with both the $q_2^{\mathrm{S}}$ and the $\epsilon_2$. Furthermore, a significant forward/backward asymmetry of $v_2(\eta)$ is observed for events selected on $q_2^{\mathrm{S}}$ but not $\epsilon_2$. This asymmetry is already observed outside the $\eta$ range covered by subevent S, but is bigger towards larger $|\eta|$. This asymmetry may reflect the dynamical fluctuations exposed by the $q_2^{\mathrm{S}}$ selection. Additional cross-checks performed by choosing subevent S in a more restricted $\eta$ range show similar asymmetry (see Fig.~\ref{fig:a0} in the Appendix).

Based on the good agreement between the three $v_2$ estimations in Fig.~\ref{fig:eta1}, they are combined into a single $v_2(\eta)$ result. Good agreement is also observed for higher harmonics, hence they are combined in the same way. The resulting $v_2(\eta)$--$v_6(\eta)$ are shown in Fig.~\ref{fig:eta2} for events with lower 10\% and upper 10\% of the values of $q_2^{\mathrm{S}}$. The asymmetry of $v_n$ in $\eta$ is much weaker for the higher-order harmonics. The values of $v_n(\eta)$ for $n>3$ are also seen to be positively correlated with $v_2$, $i.e.$ events with large $q_2^{\mathrm{S}}$ also have bigger $v_n(\eta)$. On the other hand, $v_3$ values are observed to decrease with increasing $q_2^{\mathrm{S}}$. This decrease reflects the anti-correlation between $\epsilon_2$ and $\epsilon_3$ in Fig.~\ref{fig:i1} (also confirm by ATLAS data~\cite{ATLAS2014-022}). Figure~\ref{fig:eta3} quantifies the forward/backward asymmetry of $v_2$--$v_6$ in two $\eta$ ranges: $-5<\eta<-4$ and $4<\eta<5$. Clear asymmetry can be seen for $v_2$, $v_4$ and $v_5$, but not for $v_3$. This behavior re-enforces our earlier conclusion that the correlation between $v_2$ and $v_3$ in the AMPT model is mostly geometrical, $i.e.$ reflecting correlation between $\epsilon_2$ and $\epsilon_3$.

\begin{figure}[h!]
\centering
\includegraphics[width=1\linewidth]{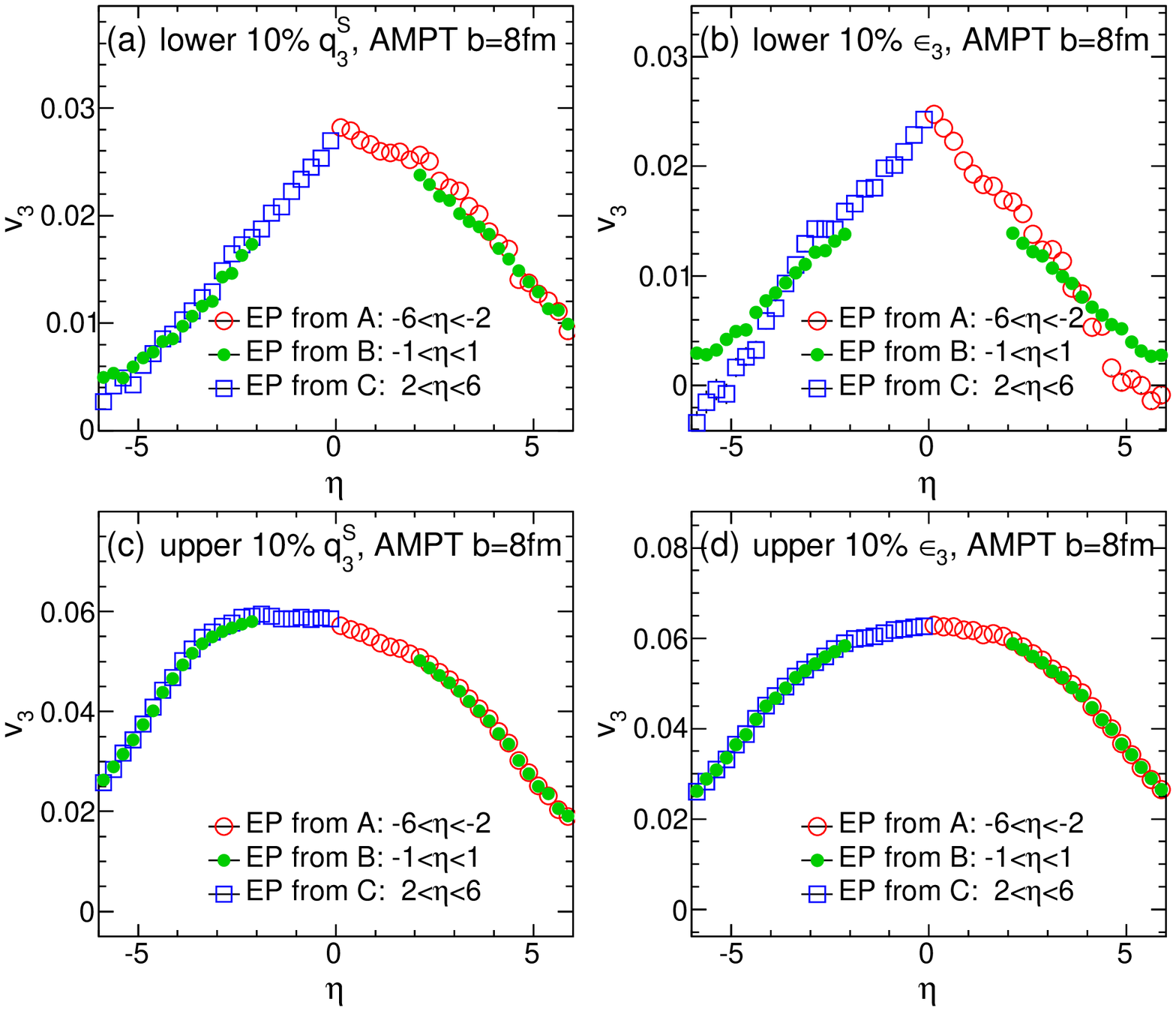}
\caption{\label{fig:eta1b} (Color online) The $v_3(\eta)$ for events selected for lower 10\% (top panels) and upper 10\% (bottom panels) of the values of either $q_3^{\mathrm{S}}$ (left panels) or $\epsilon_3$ (right panels) for AMPT Pb+Pb events with $b=8$~fm. In each case, the integral $v_3$ calculated for particles in $0.1<\pT<5$ GeV relative to the event plane of subevent A, B and C (Their $\eta$ coverages are indicated in the legend) with a minimum $\eta$ gap of 1 unit are shown. }
\end{figure}

An identical analysis is also performed for events selected on $q_3^{\mathrm{S}}$ or $\epsilon_3$. The $v_n(\eta)$ for events with the upper 10\% and lower 10\% values of $q_3^{\mathrm{S}}$ or $\epsilon_3$ are shown in Fig.~\ref{fig:eta1b}. A strong $\eta$ asymmetry is observed as a result of $q_3^{\mathrm{S}}$ selection, but not for $\epsilon_3$ selection. Nevertheless, the overall magnitude of the $v_3$ is similar between the two selections. In the $-6<\eta<-2$ range where $q_3^{\mathrm{S}}$ is calculated, $v_3(\eta)$ values for events with the lower 10\% of $q_3^{\mathrm{S}}$ drop to below zero. This implies that the $\Phi_3$ angle for large negative $\eta$ region become out of phase with the $\Phi_3$ angle in the large positive range. This $\Phi_3$ angle decorrelation is also observed for events selected with lower 10\% of $\epsilon_3$ values as shown in the top-right panel of Fig.~\ref{fig:eta1b}. This behavior suggests that in the AMPT model, rapidity decorrelation of $v_3$ is stronger for events with small $\epsilon_3$ and grows towards large $|\eta|$ (negative $v_3$ implies its phase is opposite to that in the $\eta$ region used to obtain the event plane). An earlier study~\cite{Xiao:2012uw} has show evidences of $\eta$ decorrelation of $v_3$ in the AMPT; Our later studies published in separate papers trace this decorrelation to the independent fluctuations of the $\epsilon_n$ for the projectile nucleus and the $\epsilon_n$ for the target nucleus~\cite{Jia:2014vja,Jia:2014ysa}.

\begin{figure}[h!]
\centering
\includegraphics[width=1\linewidth]{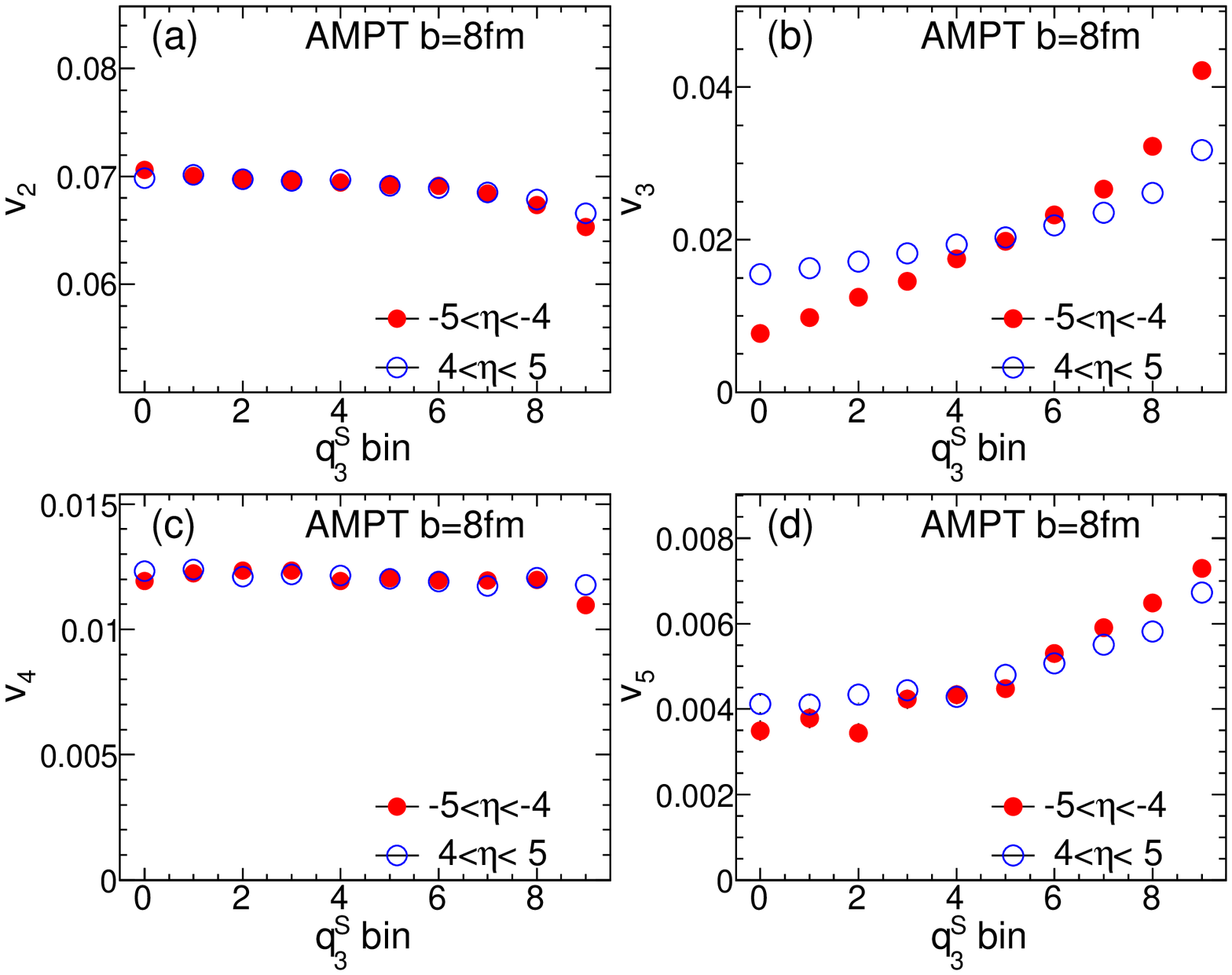}
\caption{\label{fig:eta3b} (Color online)  Dependence of the $v_n$ on $q_3^{\mathrm{S}}$ in a forward ($-5<\eta<-4$) and a backward ($4<\eta<5$) pseudorapidity ranges.}
\end{figure}

 Figure~\ref{fig:eta3b} quantifies the rapidity asymmetry of $v_n$ between $-5<\eta<-4$ and  $4<\eta<5$ as a function of $q_3^{\mathrm{S}}$ and $\epsilon_3$. The even harmonics $v_2$ and $v_4$ show little asymmetry and are nearly independent of $q_3^{\mathrm{S}}$. In contrast, the $v_5$ values show a strong $\eta$-asymmetry similar to that for $v_3$. 

\begin{figure}[h!]
\centering
\includegraphics[width=0.5\linewidth]{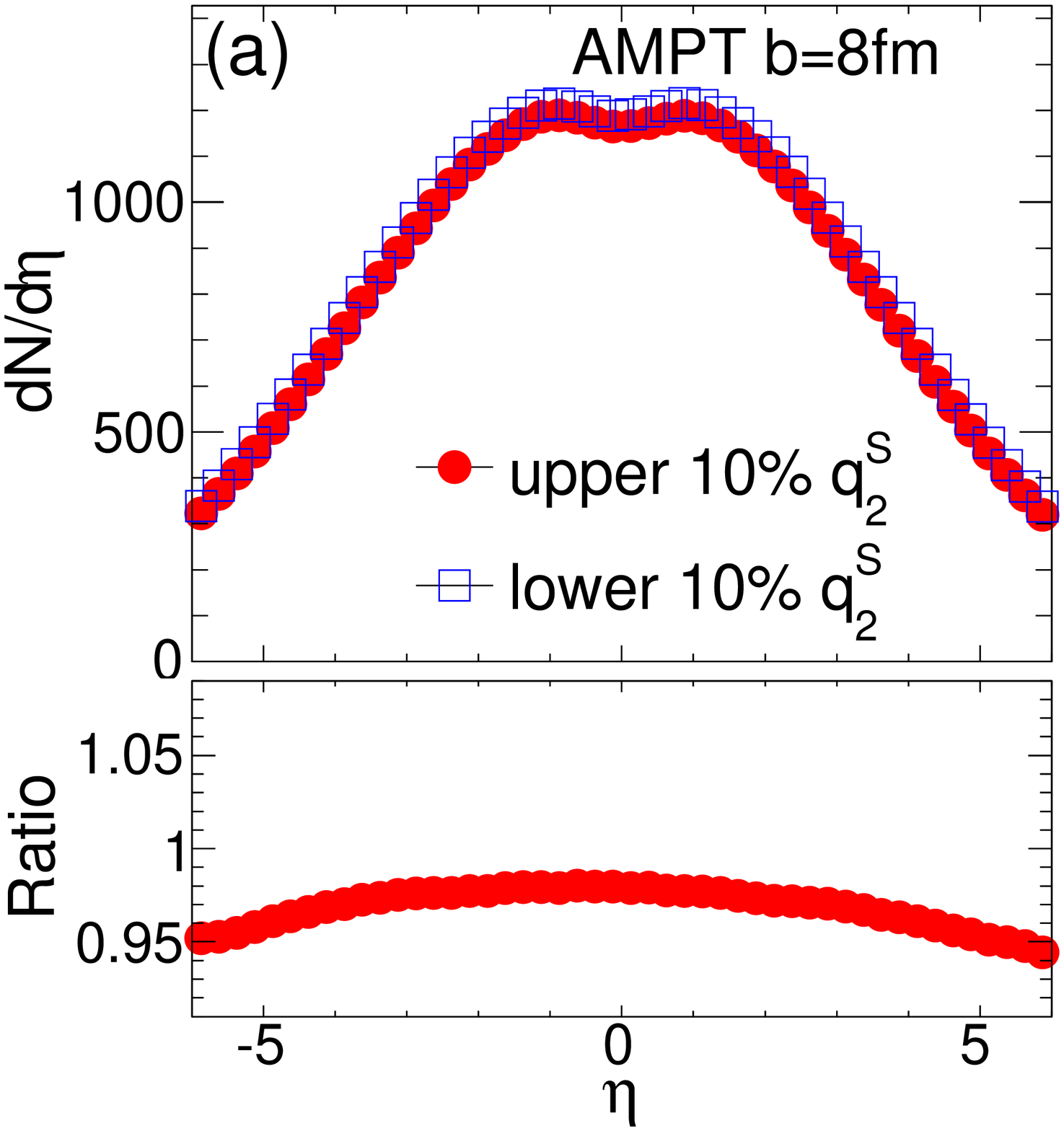}\includegraphics[width=0.5\linewidth]{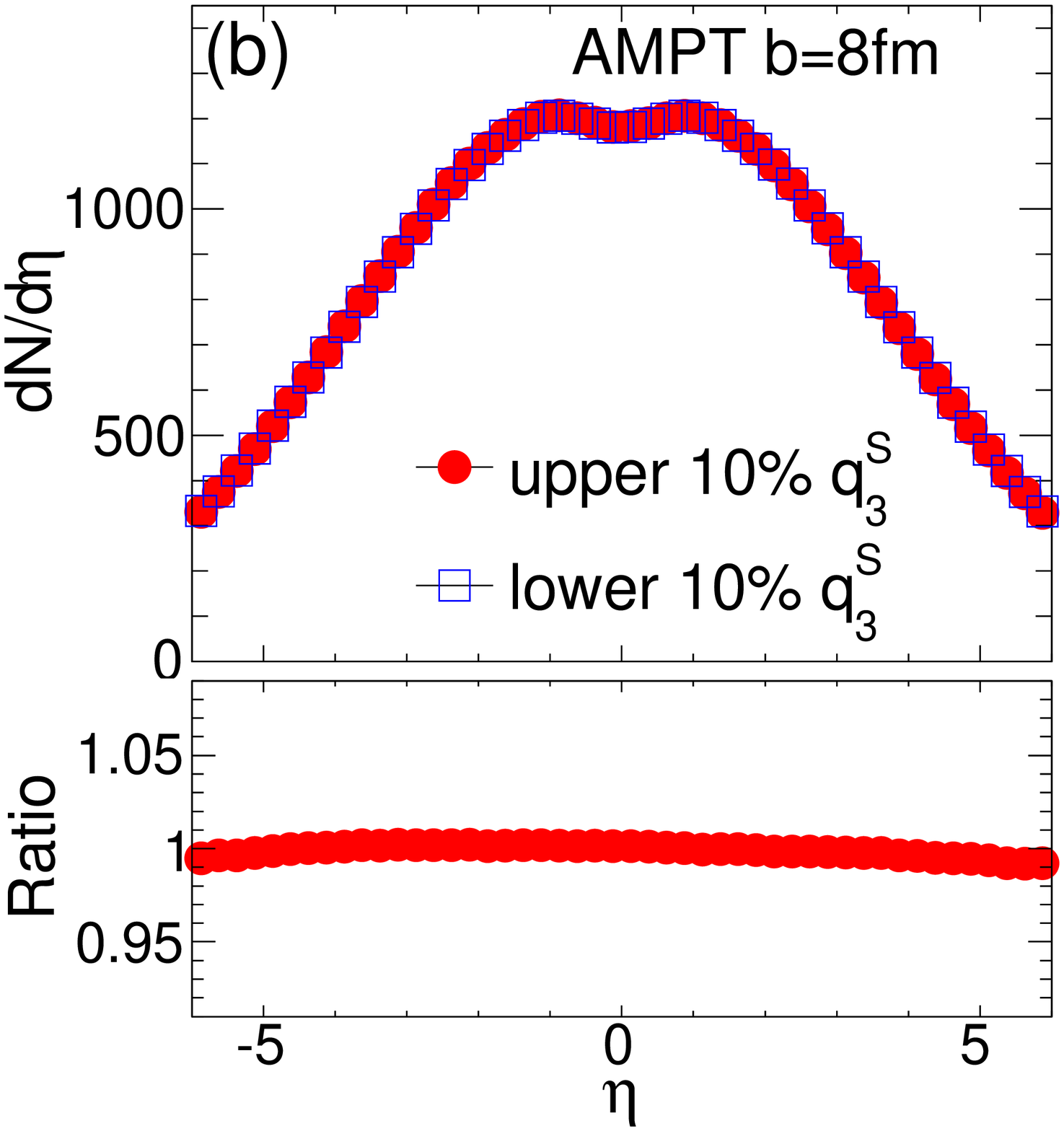}
\caption{\label{fig:eta5} (Color online) The $dN/d\eta$ distributions of all particles for events selected on $q_2^{\mathrm{S}}$ (left) and $q_3^{\mathrm{S}}$ (right).}
\end{figure}

Figure~\ref{fig:eta5} shows the particle multiplicity distributions $dN/d\eta$ for events selected on $q_2^{\mathrm{S}}$ (left) or $q_3^{\mathrm{S}}$ (right). The distributions remain largely symmetric in $\eta$ and the overall magnitude is nearly independent of the event selection. We also verified explicitly that the number of participating nucleons for the projectile and target are nearly equal for all $q_2^{\mathrm{S}}$ or $q_3^{\mathrm{S}}$ bins. This suggests that the underlying mechanism is not due to the EbyE fluctuations of the $dN/d\eta$ distribution.

\section{event-plane correlations}
\label{sec:4}
The AMPT model has been shown to reproduce~\cite{Bhalerao:2013ina} the centrality dependence of various two-plane and three-plane correlations measured by the ATLAS Collaboration~\cite{Jia:2012sa}. Here we use AMPT model to study how these correlators change with $q_n^{\mathrm{S}}$ or $\epsilon_n$. In this analysis, the two-plane correlators $\langle\cos k(\Phi_{n}-\Phi_{m})\rangle$ are calculated by correlating the EP angles from subevent A and subevent C. Each subevent provides its own estimation of the EPs, leading to two statistically independent estimates of the correlator: Type1 $\langle\cos k(\Phi_{n}^{\mathrm{A}}-\Phi_{m}^{\mathrm{C}})\rangle$ and Type2 $\langle\cos k(\Phi_{n}^{\mathrm{C}}-\Phi_{m}^{\mathrm{A}})\rangle$. The two estimates are identical for events selected on $\epsilon_2$, and hence they are averaged to obtain the final result. But for events selected based on $q_2^{\mathrm{S}}$, the two estimates can differ quite significantly. 

\begin{figure}[h!]
\centering
\includegraphics[width=1\linewidth]{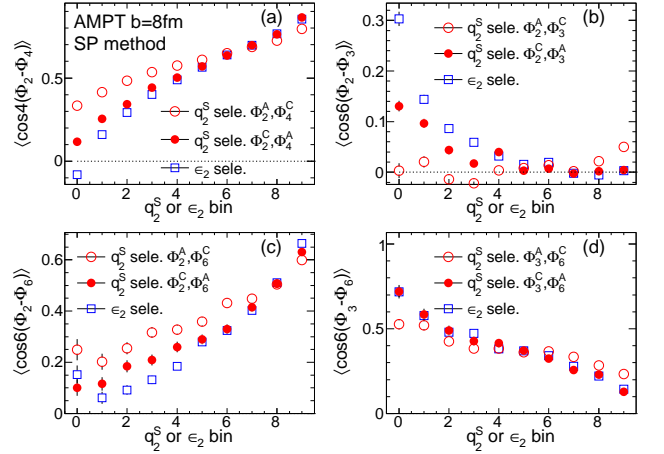}
\caption{\label{fig:ep1} (Color online) The four two-plane correlations as a function of bins of $q_2^{\mathrm{S}}$ or $\epsilon_2$ for AMPT Pb+Pb events with $b=8$~fm. The two event planes $\Phi_n$ and $\Phi_m$ are measured by subevent A and subevent C. The $q_2^{\mathrm{S}}$-binned results are presented separately for the two combinations: $\langle\cos k(\Phi_{n}^{\mathrm{A}}-\Phi_{m}^{\mathrm{C}})\rangle$ (open circles) and $\langle\cos k(\Phi_{n}^{\mathrm{C}}-\Phi_{m}^{\mathrm{A}})\rangle$ (solid circles).}
\end{figure}

Figure~\ref{fig:ep1} shows the values of four two-plane correlators in bins of $q_2^{\mathrm{S}}$ or $\epsilon_2$. The values of the correlators are observed to increase strongly with increasing $q_2^{\mathrm{S}}$ or $\epsilon_2$. The two estimates based on $q_2^{\mathrm{S}}$ selection differ significantly, reflecting the influence of longitudinal flow fluctuations exposed by the $q_2^{\mathrm{S}}$ selection. Interestingly, the correlators whose $\Phi_2$ angle is calculated in subevent C agree very well with those based on $\epsilon_2$ event shape selection, such as $\langle\cos 4(\Phi_{2}^{\mathrm{C}}-\Phi_{4}^{\mathrm{A}})\rangle$. This is because $\Phi_{2}^{\mathrm{C}}$ is expected to be less dependent on the $q_2^{\mathrm{S}}$ selection than $\Phi_{2}^{\mathrm{A}}$ (see Fig.~\ref{fig:eta3}(a)). These observations suggest that the dependence of $\langle\cos 4(\Phi_{2}^{\mathrm{C}}-\Phi_{4}^{\mathrm{A}})\rangle$ and $\langle\cos 6(\Phi_{2}^{\mathrm{C}}-\Phi_{6}^{\mathrm{A}})\rangle$ on $q_2^{\mathrm{S}}$ reflects mainly the change in the initial geometry and the ensuing non-linear effects in the final state. Note that the last bin in each panel represents the value obtained without event shape selection, which agrees between the three calculations by construction.

\begin{figure}[h!]
\centering
\includegraphics[width=1\linewidth]{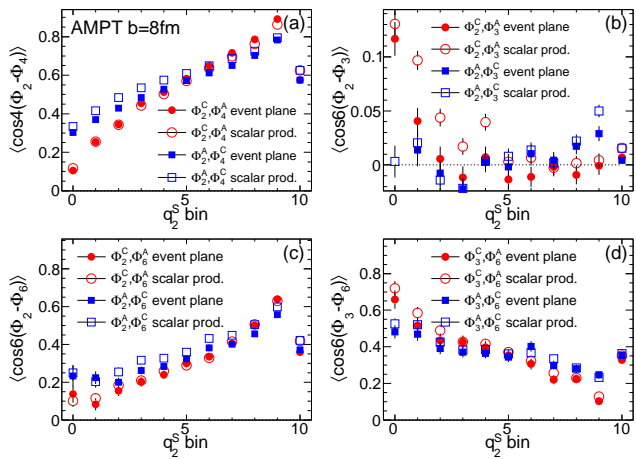}
\caption{\label{fig:ep2} (Color online) The four two-plane correlators in bins of $q_2^{\mathrm{S}}$ for AMPT Pb+Pb events with $b=8$~fm, shown for two different combinations of the event-plane angles (solid symbols), and compared with the correlations calculated via the scalar product method (open symbols).}
\end{figure}

Figure~\ref{fig:ep2} compares various two-plane correlators calculated via the EP method and the SP method given by Eqs.~\ref{eq:m4a}-\ref{eq:m5}. The SP method is observed to give systematically higher values for Type1 correlators where the first angle is measured by subevent A, while it gives consistent or slightly lower values for Type2 correlators. The last bin in each panel shows the result obtained without event shape selection, where the values from the SP method are always higher, as expected~\cite{Bhalerao:2013ina}. 

\begin{figure}[h!]
\centering
\includegraphics[width=1\linewidth]{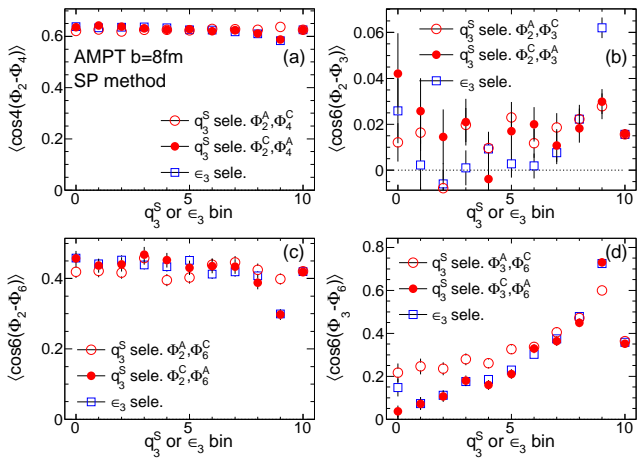}
\caption{\label{fig:ep1b} (Color online) Two two-plane correlators as a function of either $q_3^{\mathrm{S}}$ or $\epsilon_3$ for AMPT Pb+Pb events with $b=8$~fm. The event planes $\Phi_n$ and $\Phi_m$ are measured by subevent A and subevent C. The $q_3^{\mathrm{S}}$-binned results are presented separately for the two estimates: $\langle\cos k(\Phi_{n}^{\mathrm{A}}-\Phi_{m}^{\mathrm{C}})\rangle$ (open circles) and $\langle\cos k(\Phi_{n}^{\mathrm{C}}-\Phi_{m}^{\mathrm{A}})\rangle$ (solid circles).}
\end{figure}

Figure~\ref{fig:ep1b} shows $\langle\cos 6(\Phi_{2}-\Phi_{3})\rangle$ and $\langle\cos 6(\Phi_{3}-\Phi_{6})\rangle$ in bins of $q_3^{\mathrm{S}}$ or $\epsilon_3$. The first correlator shows little dependence on $q_3^{\mathrm{S}}$ or $\epsilon_3$, while the second correlator does. This is in sharp contrast to the results seen in Fig.~\ref{fig:ep1}, where both correlators show strong but opposite dependence on $q_2^{\mathrm{S}}$ or $\epsilon_2$. This behavior is consistent with a strong coupling between $v_6$ and $v_2$, $v_6$ and $v_3$, but weak coupling between $v_2$ and $v_3$.

To calculate three-plane correlations, subevents A, B and C are used. Each subevent provides its own estimation of the three EP angles, and hence there are $3!=6$ independent ways of estimating a given three-plane correlator. For $c_nn\Phi_{n}+c_mm\Phi_{m}+c_ll\Phi_{l}$ with $n<m<l$, these six estimates are labelled as the following:
\begin{itemize}
\item Type 1a: $c_nn\Phi_{n}^{\mathrm{B}}+c_mm\Phi_{m}^{\mathrm{A}}+c_ll\Phi_{l}^{\mathrm{C}}$
\item Type 1b: $c_nn\Phi_{n}^{\mathrm{B}}+c_mm\Phi_{m}^{\mathrm{C}}+c_ll\Phi_{l}^{\mathrm{A}}$
\item Type 2a: $c_nn\Phi_{n}^{\mathrm{A}}+c_mm\Phi_{m}^{\mathrm{B}}+c_ll\Phi_{l}^{\mathrm{C}}$
\item Type 2b: $c_nn\Phi_{n}^{\mathrm{C}}+c_mm\Phi_{m}^{\mathrm{B}}+c_ll\Phi_{l}^{\mathrm{A}}$
\item Type 3a: $c_nn\Phi_{n}^{\mathrm{A}}+c_mm\Phi_{m}^{\mathrm{C}}+c_ll\Phi_{l}^{\mathrm{B}}$
\item Type 3b: $c_nn\Phi_{n}^{\mathrm{C}}+c_mm\Phi_{m}^{\mathrm{A}}+c_ll\Phi_{l}^{\mathrm{B}}$
\end{itemize}
For events selected on $\epsilon_n$, the symmetry of the $\eta$-coverage between A and C reduces them into three equivalent pairs of estimates. However for events selected on $q_n^{\mathrm{S}}$, all six estimates can be different.

\begin{figure*}[t!]
\centering
\includegraphics[width=1\linewidth]{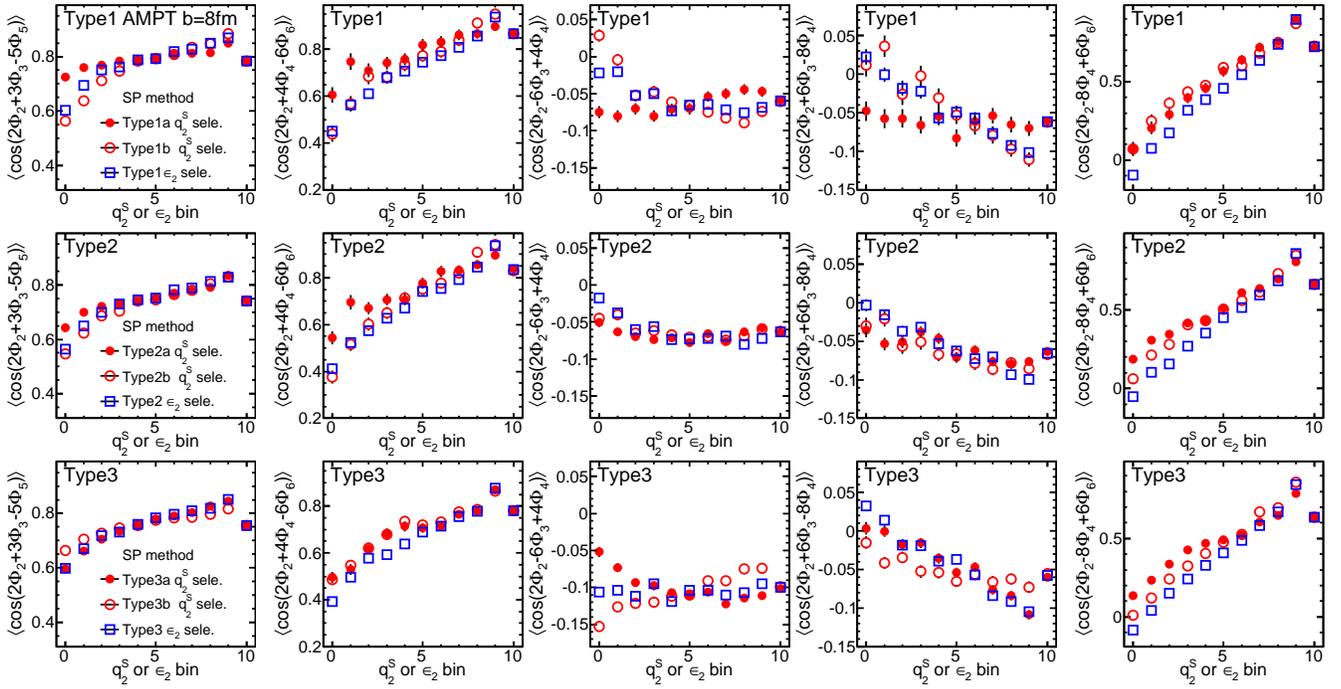}
\caption{\label{fig:ep3} (Color online) 
The five three-plane correlators (from left to right) as a function of either $q_2^{\mathrm{S}}$ (solid symbols) or $\epsilon_2$ (open symbols) for AMPT Pb+Pb events with $b=8$~fm. The results for the three type groups are shown in different row: Type1a and Type1b (top row), Type2a and Type2b (middle row), Type2a and Type2b (bottom row). The last bin in each panel shows the inclusive result.}
\end{figure*}

Figure~\ref{fig:ep3} summarizes the results for five three-plane correlators, one for each column, calculated for events classified by $q_2^{\mathrm{S}}$ or $\epsilon_2$~\footnote{We also calculate these correlators using the SP method (a.la. $q_n$-weights), the dependences on $q_2^{\mathrm{S}}$ are found to be qualitatively the same, though the overall magnitudes may differ (see Fig.~\ref{fig:a3}).}, including a previously unnoticed strong correlator $\left\langle\cos (2\Phi_{2}-8\Phi_{4}+6\Phi_6)\right\rangle$. The six estimates of each correlator are grouped into three pairs and are shown in the three rows. Many of these correlators exhibit a breaking of the symmetry between subevent A and C, see for example the Type1 and Type2 for the first two correlators, as well as the Type1 and Type3 for the third and fourth correlators. Some correlators even show opposite dependence on $q_2^{\mathrm{S}}$, such as the Type1 and Type3 for the fourth correlator ($\left\langle\cos (2\Phi_{2}-6\Phi_{3}+4\Phi_4)\right\rangle$). In most cases, however, the overall dependences on $\epsilon_2$ bin (open circles) are reasonably captured by the dependence on $q_2^{\mathrm{S}}$ (open boxes), implying that these correlations reflect mainly the intrinsic hydrodynamic response to the change in the selected initial geometry. 

\begin{figure*}[!t]
\centering
\includegraphics[width=1\linewidth]{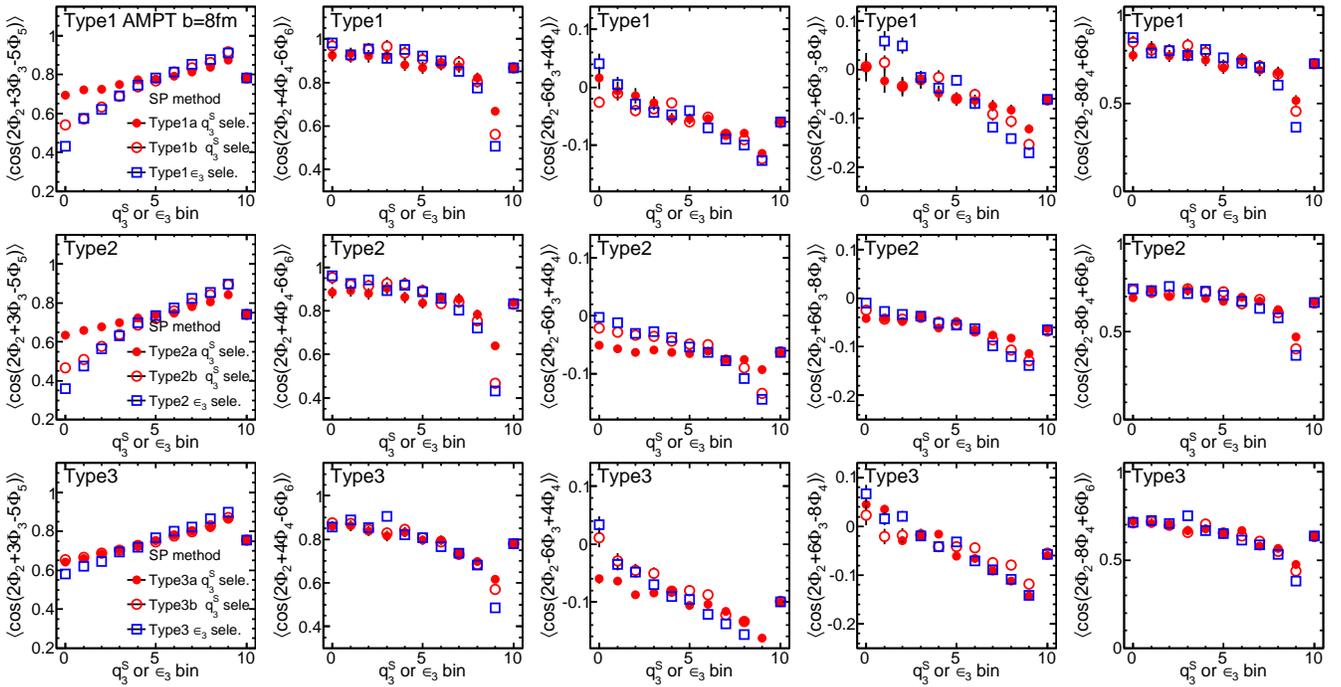}
\caption{\label{fig:ep3b} (Color online) 
The five three-plane correlators (from left to right) as a function of either $q_3^{\mathrm{S}}$ (solid symbols) or $\epsilon_3$ (open symbols) for AMPT Pb+Pb events with $b=8$~fm. The results for the three type groups are shown in different row: Type1a and Type1b (top row), Type2a and Type2b (middle row), Type2a and Type2b (bottom row). The last bin in each panel shows the inclusive result.}
\end{figure*}

Figure~\ref{fig:ep3b} summarizes the results for the same five three-plane correlators, but calculated for events classified by $q_3^{\mathrm{S}}$ or $\epsilon_3$. Comparing with Fig.~\ref{fig:ep3}, we find that the values of $\left\langle\cos (2\Phi_{2}-3\Phi_{3}+5\Phi_5)\right\rangle$ are more sensitive to $q_3^{\mathrm{S}}$ or $\epsilon_3$ than to $q_2^{\mathrm{S}}$ or $\epsilon_2$ for small bin numbers, possibly reflecting stronger rapidity decorrelation effects associated with $\Phi_3$ seen for small $q_3^{\mathrm{S}}$ or $\epsilon_3$ in Fig.~\ref{fig:eta1b}. The correlators $\left\langle\cos (2\Phi_{2}-6\Phi_{3}+4\Phi_4)\right\rangle$ and $\left\langle\cos (2\Phi_{2}+6\Phi_{3}-8\Phi_4)\right\rangle$ show better agreement between the six types than when they are selected by $q_2^{\mathrm{S}}$ or $\epsilon_2$ as shown in Fig.~\ref{fig:ep3}, again reflecting the weak correlation between $v_3$ and $v_2$, and between $v_3$ and $v_4$.

\section{Discussion and Summary}
\label{sec:5}

This paper studies two sets of flow observables involving correlations between harmonic flow coefficients $v_n$ and their phases $\Phi_n$, utilizing the recently proposed event shape selection technique~\cite{Schukraft:2012ah}. The shape of the collision geometry is selected by cutting on $q_n$ (the magnitude of the flow vector in a subevent) and eccentricity $\epsilon_n$. The $p(v_m)$ or the differential distribution of event-averaged $v_m$, and the event-plane correlations $p(\Phi_m, \Phi_l, ....)$ are then be studied in each $q_n$ or $\epsilon_n$ bin. A special case of interest is the $v_n(\eta)$ for events selected on $q_n$, which is sensitive to rapidity fluctuations of collective flow. The feasibility of measuring these new observables is investigated using the AMPT model. This model combines the Glauber initial fluctuating geometry with collective flow generated by partonic transport, and hence allows one to correlate the initial geometry information, $e.g.$ ($\epsilon_n$, $\Phi_n^*$), with ($v_n$,$\Phi_n$) on a EbyE basis. Since the AMPT model describes reasonably the experimentally measured $\pT$ spectra, $v_n$~\cite{Xu:2011fe,Xu:2011fi}, and event-plane correlations~\cite{Bhalerao:2013ina}, it should provide a good benchmark for the performance of event shape selection, as well as provide qualitative understanding of the physics behind these new observables.

To summarize, our study has been performed for Pb+Pb collisions at $\sqrt{s_{NN}}=2.76$ TeV with fixed impact parameter $b=8$~fm, which corresponds to $\sim30\%$ centrality. The event shape selection is performed on the $q_2^{\mathrm{S}}$ and $q_3^{\mathrm{S}}$ calculated using half of the particles in $-6<\eta<-2$, as well as on $\epsilon_2$ and $\epsilon_3$. The EP angles used to calculate the EP correlators are calculated in subevent A (the other half particles in $-6<\eta<-2$), subevent B covering $-1<\eta<1$ and subevent C covering $2<\eta<6$. Our main findings can be summarized below:
\begin{itemize}
\item
The eccentricity $\epsilon_n$ distribution is very broad for events with fixed impact parameter and correlates well with $q_n$ for $n=2$ and $3$, hence events with different ellipticity or triangularity can be selected precisely using $q_2$ and $q_3$, respectively.
\item
The participant plane correlations, such as $\langle\cos4(\Phi_{2}^*-\Phi_{4}^*)\rangle$, $\langle\cos6(\Phi_{2}^*-\Phi_{6}^*)\rangle$ and $\langle\cos6(\Phi_{3}^*-\Phi_{6}^*)\rangle$, are observed to vary strongly with $\epsilon_2$ or $\epsilon_3$, and much of these dependences are preserved when selecting on $q_2$ or $q_3$. 
\item
Significant correlations or anti-correlations are observed among the $\epsilon_n$, such as $p(\epsilon_2, \epsilon_3)$, $p(\epsilon_1, \epsilon_3)$ and $p(\epsilon_2, \epsilon_4)$. These correlations can be easily probed by the event shape selection technique, some of these correlations, $e.g.$ $p(\epsilon_1, \epsilon_3)$ and $p(\epsilon_2, \epsilon_3)$, are expected to survive to the final state. Indeed, the correlation between $\epsilon_2$ and $\epsilon_3$ seems to be captured by the observed correlation between $v_2$ and $v_3$ in the AMPT model. 
\item
The overall $v_n$ values in each $q_n^{\mathrm{S}}$ bin are similar to that in the corresponding $\epsilon_n$ bin ($n=2$ and 3). However a strong forward/backward asymmetry of $v_n(\eta)$ is observed in $q_n^{\mathrm{S}}$ selected events (also $v_4$, $v_5$ for $q_2^{\mathrm{S}}$ and $v_5$ for $q_3^{\mathrm{S}}$), reflecting the dynamical fluctuations either in the initial state~\cite{Pang:2012he,Pang:2013pma} or during the collective expansion present in the AMPT model. These dynamical fluctuations are exposed by the $q_n^{\mathrm{S}}$ selection, probably because they are either short range in $\eta$ or have strong $\eta$ dependence. These dynamical effects also contribute strongly to event-plane decorrelations. We also observe that the $dN/d\eta$ distribution is not affected much by the event shape selection.
\item
The $v_n$ values are always positively correlated with $q_n^{\mathrm{S}}$, however other harmonics $v_m$ for $m\neq n$ show non-trivial dependence on $q_n^{\mathrm{S}}$. For $q_2^{\mathrm{S}}$ selection, $v_4$, $v_5$ and $v_6$ show positive correlation, while $v_3$ shows negative correlation consistent with the behavior of $p(\epsilon_2, \epsilon_3)$. For $q_3^{\mathrm{S}}$ selection, $v_5$ shows positive correlation, while the $v_2$ and $v_4$ show slight negative correlation. These correlations are qualitatively consistent with couplings between different harmonics expected for collective flow~\cite{Teaney:2012ke}.
\item
The strengths of several event-plane correlators vary strongly with $q_n^{\mathrm{S}}$ and $\epsilon_n$. These variations are observed to be similar between the EP method and the SP method. In many cases, they are also found to be similar between $q_n^{\mathrm{S}}$ selection and $\epsilon_n$ selection, namely the $\langle\cos 4(\Phi_{2}^{\mathrm{C}}-\Phi_{4}^{\mathrm{A}})\rangle$, $\langle\cos 6(\Phi_{2}^{\mathrm{C}}-\Phi_{6}^{\mathrm{A}})\rangle$, and many of $\langle\cos (2\Phi_{2}+3\Phi_3-5\Phi_{5})\rangle$, $\langle\cos (2\Phi_{2}+4\Phi_4-6\Phi_{6})\rangle$ and $\langle\cos (2\Phi_{2}-8\Phi_4+6\Phi_{6})\rangle$. For these correlators, the change of signal reflects mainly the selection on the collision geometry. In some other cases, results differ significant between $q_n^{\mathrm{S}}$ and $\epsilon_n$ selections, such as $\langle\cos 4(\Phi_{2}^{\mathrm{A}}-\Phi_{4}^{\mathrm{C}})\rangle$, $\langle\cos 6(\Phi_{2}^{\mathrm{A}}-\Phi_{6}^{\mathrm{C}})\rangle$ and $\langle\cos (2\Phi_{2}^{\mathrm{B}}+3\Phi_3^{\mathrm{A}}-5\Phi_{5}^{\mathrm{C}})\rangle$ between $q_2^{\mathrm{S}}$ and $\epsilon_2$, and $\langle\cos (2\Phi_{2}^{\mathrm{B}}+3\Phi_3^{\mathrm{A}}-5\Phi_{5}^{\mathrm{C}})\rangle$ and $\langle\cos (2\Phi_{2}^{\mathrm{A}}+3\Phi_3^{\mathrm{B}}-5\Phi_{5}^{\mathrm{C}})\rangle$ between $q_3^{\mathrm{S}}$ and $\epsilon_3$. For these correlators, the change of signal is probably also affected by the dynamical fluctuations present in the AMPT model.
\item
For some event-plane correlators, the signal for different types show very different dependence on $q_2^{\mathrm{S}}$. For example $\langle\cos (2\Phi_{2}^{\mathrm{A}}-6\Phi_3^{\mathrm{C}}+4\Phi_{4}^{\mathrm{B}})\rangle$ and $\langle\cos (2\Phi_{2}^{\mathrm{C}}-6\Phi_3^{\mathrm{A}}+4\Phi_{4}^{\mathrm{B}})\rangle$ have opposite dependence, and both differ from the dependence on $\epsilon_2$. This may be a result of the rapidity fluctuations of $v_3(\eta)$, as shown in Fig.~\ref{fig:eta1b}.
\end{itemize}

Heavy ion collisions at RHIC and LHC are a finite number system. A typical Pb+Pb collision involves a few hundred colliding nucleons and produces on the order of 10,000 particles in the final state. Both the collision geometry defined by the nucleons and the collective expansion of produced particles can fluctuate strongly event-by-event. It remains a challenge to disentangle the fluctuations at the initial state and dynamical fluctuations generated in the collective expansion, a prerequisite for understanding the physics behind these fluctuations. Future progress requires a detailed and systematic study of the correlations between all $v_n$ and their phases $\Phi_n$: $p(v_n,v_m,..., \Phi_n, \Phi_m,...)$, in order to disentangle different sources of fluctuations. The study presented in this paper attempts to establish the methodology for accessing these correlations and provides the initial guidance on how these correlations are connected to the initial geometry and dynamics in the collective expansion. Much more theoretical efforts, especially those based on the application of event shape selection technique in EbyE hydrodynamics, will provide more realistic and detailed insights on the nature of the fluctuations and non-linear dynamics in various stages of the heavy ion collisions.

This research is supported by NSF under grant number PHY-1305037 and by DOE through BNL under grant number DE-AC02-98CH10886.
\bibliography{eventshapev4}{}
\bibliographystyle{apsrev4-1}

\appendix

\section{Additional figures}
\label{sec:7}

\begin{figure}[h!]
\centering
\includegraphics[width=1\linewidth]{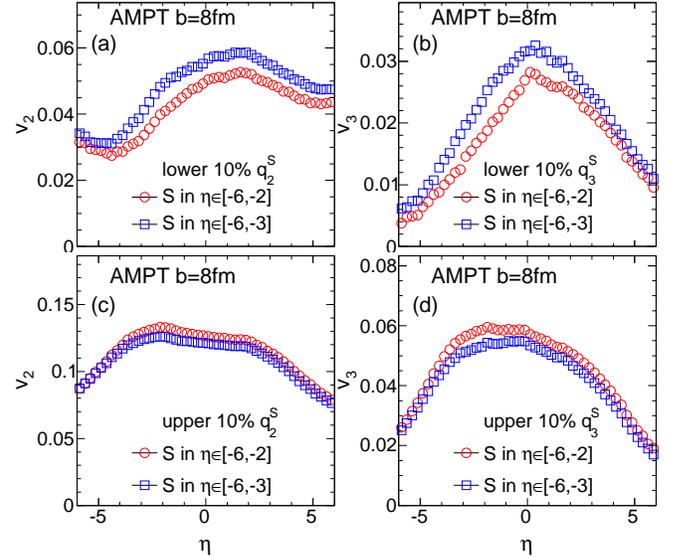}
\caption{\label{fig:a0} The $v_2(\eta)$ (left) and $v_3(\eta)$ (right) selected for lower 10\% (top) and upper 10\% (bottom) of the $q_n^{S}$ values. The results are shown for the subevent S (used for event shape selection) defined in two $\eta$ ranges: $-6<\eta<-2$ (circles) and $-6<\eta<-3$ (boxes).}
\end{figure}

Figure~\ref{fig:a0} compares the $v_n(\eta)$ obtained for lower 10\% (top) and upper 10\% (bottom) of the $q_n^{S}$ values and for two different $\eta$ ranges for the subevent S: $-6<\eta<-2$ (circles) and $-6<\eta<-3$ (boxes). The event shape selection is less effective when subevent S has a smaller $\eta$ range, but the overall $\eta$ asymmetry is similar between the two cases.

\begin{figure*}[h!]
\centering
\includegraphics[width=0.75\linewidth]{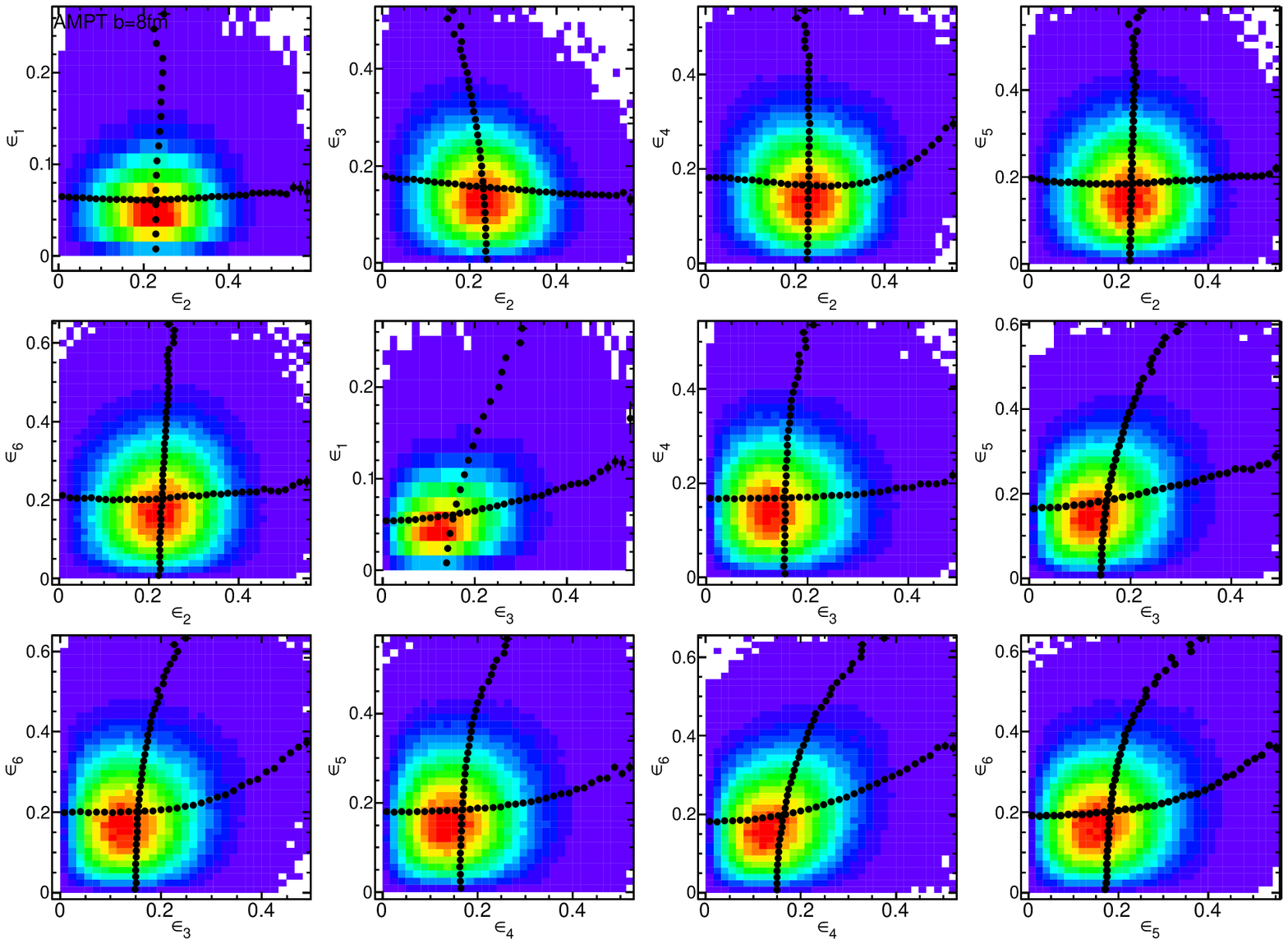}
\caption{\label{fig:a1} (Color online)  Selected correlations between $\epsilon_n$ of different order for Pb+Pb events at $b=8$~fm.}
\end{figure*}
\begin{figure*}[!h]
\centering
\includegraphics[width=0.75\linewidth]{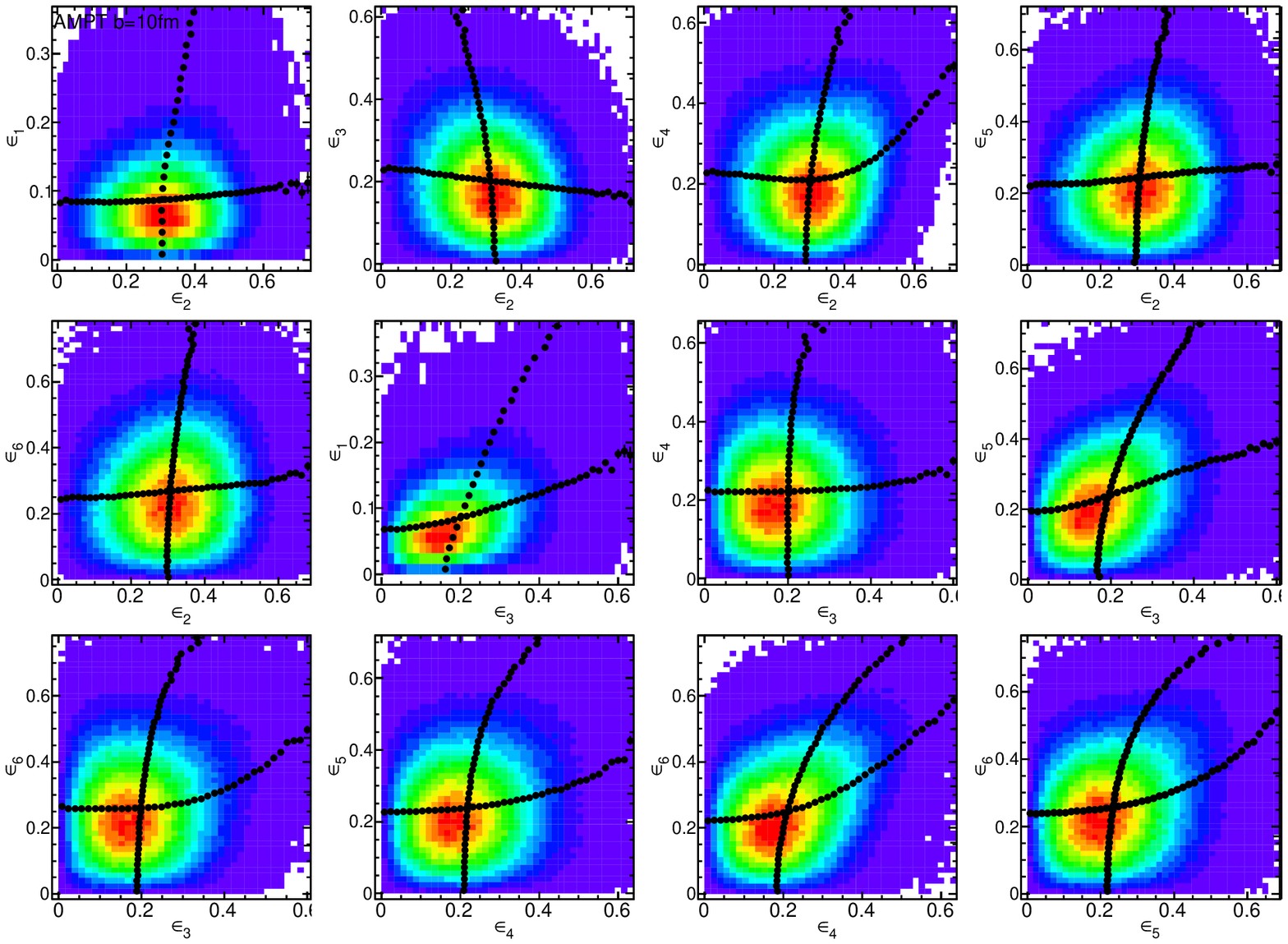}
\caption{\label{fig:a2} (Color online)  Selected correlations between $\epsilon_n$ of different order for Pb+Pb events at $b=10$~fm.}
\end{figure*}
\begin{figure*}[!t]
\centering
\includegraphics[width=1\linewidth]{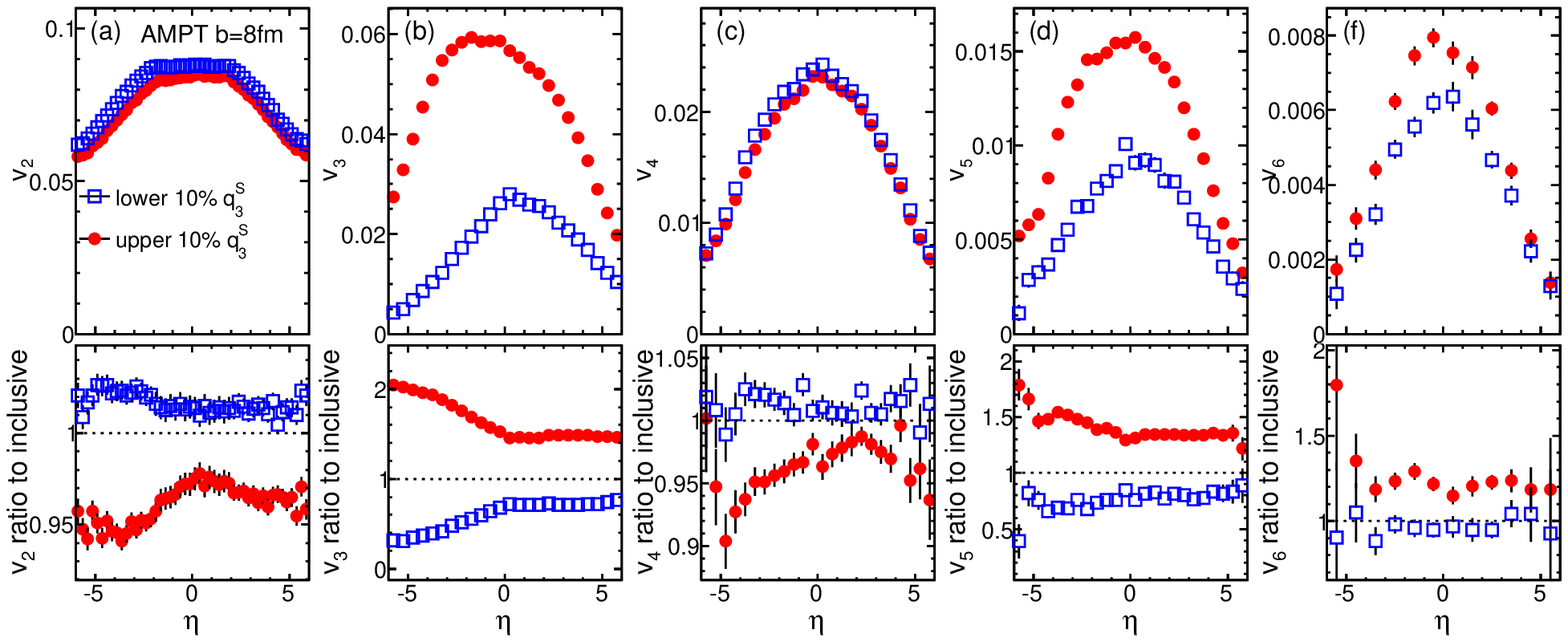}
\caption{\label{fig:aeta2} (Color online) $v_n(\eta)$ for events selected for lower 10\% (open symbols) and upper 10\% (solid symbols) of the values of $q_3^{\mathrm{S}}$ for AMPT Pb+Pb events with $b=8$~fm. Results are shown for $v_2(\eta)$, $v_3(\eta)$,..., $v_6(\eta)$ from left panel to the right panel. The ratios of $v_n(\eta)$ beween events with $q_2^{\mathrm{S}}$ selection to the inclusive events are shown in the bottom panels.}
\end{figure*}
\begin{figure*}[!t]
\centering
\includegraphics[width=1\linewidth]{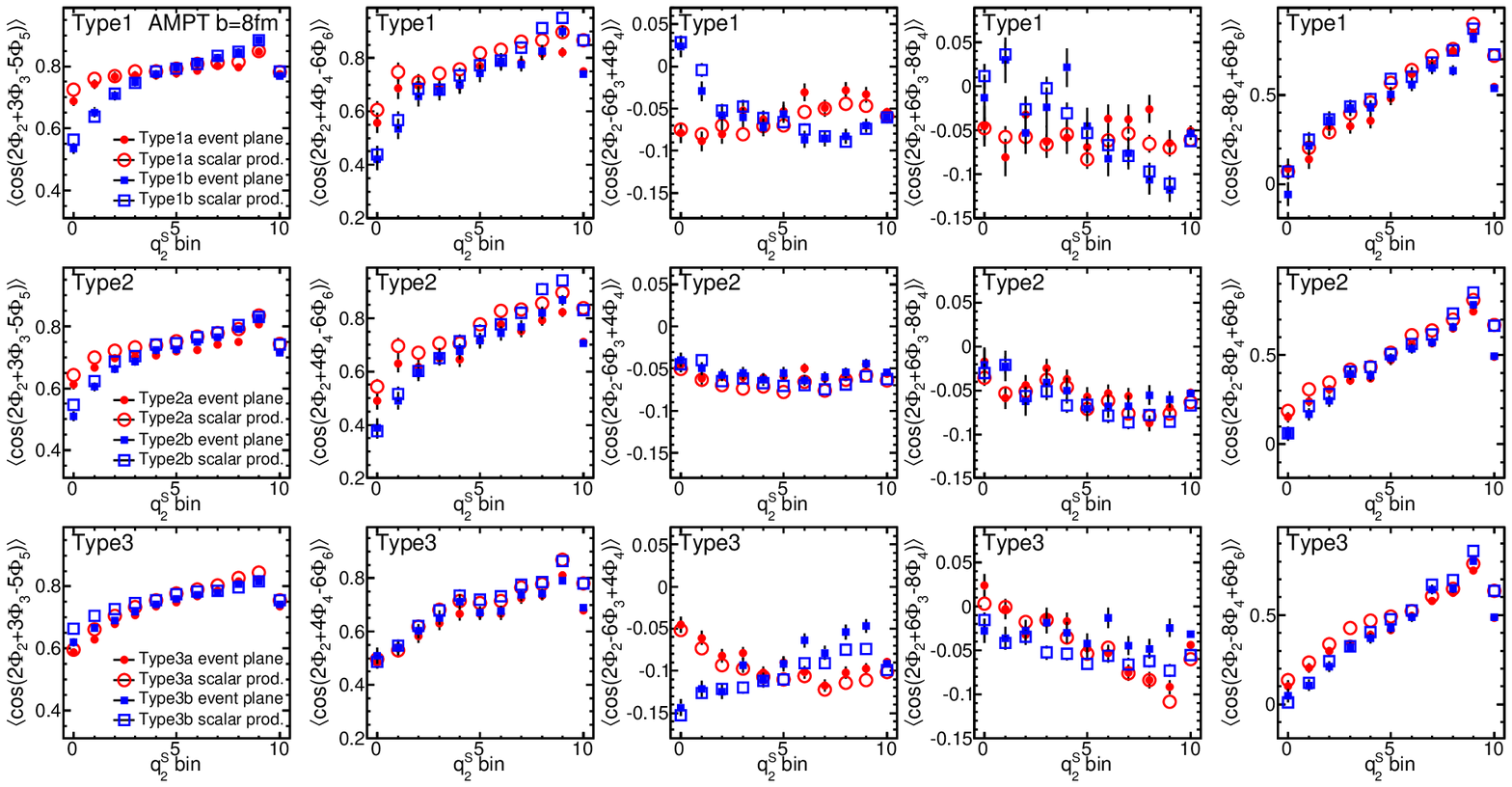}
\caption{\label{fig:a3} (Color online) The five three-plane correlators (from left to right) as a function of $q_2^{\mathrm{S}}$ event class for AMPT Pb+Pb events with $b=8$~fm. The results for the three type groups are shown in different row: Type1a and Type1b (top row), Type2a and Type2b (middle row), Type2a and Type2b (bottom row) and they are compared between the EP method and the SP method. The last bin in each panel shows the result for inclusive events.}
\end{figure*}


For completeness, Figs.~\ref{fig:a1} and \ref{fig:a2} show various correlations between $\epsilon_n$ and $\epsilon_m$ for AMPT Pb+Pb events with two fixed impact parameters, respectively. Many types of correlations, in additional to those discussed in Section~\ref{sec:2} can be identified. Figure~\ref{fig:aeta2} shows the $\eta$ dependence of the $v_n(\eta)$ for events selected based on the $q_3^{\mathrm{S}}$. It shows clearly that the forward/backward asymmetry of $v_5(\eta)$ is strongly correlated with that of the $v_3(\eta)$. The $v_5(\eta)$ also exhibits $\eta$-asymmetry similar to that observed for $v_3(\eta)$. Figures~\ref{fig:a3} shows various three-plane correlators compared between the event-plane method and the scalar product method; the dependences on $q_2^{\mathrm{S}}$ are similar between the two methods.
\end{document}